\documentclass[%
 aip,
 amsmath,amssymb,
]{revtex4-1}

\usepackage[ruled, linesnumbered]{algorithm2e}
\usepackage{algorithmicx}
\usepackage{algcompatible}
\usepackage{algpseudocode}
\usepackage{amsmath}
\usepackage{bigints}  
\usepackage{bm}
\usepackage{color}
\usepackage{etoolbox}
\usepackage{float}
\usepackage[T1]{fontenc}
\usepackage{graphicx}
\usepackage[utf8]{inputenc}
\usepackage{mathptmx}
\usepackage{multirow}
\usepackage{stmaryrd}  
\usepackage[caption = false]{subfig}

\makeatother
\begin{document}

\preprint{AIP/123-QED}

\title[Macroscopic data extraction from microscopic fields]{A method to extract macroscopic interface data from microscale rough/porous wall flow fields}
\author{Vedanth N Kuchibhotla}
\affiliation{ 
School of Mechanical Sciences, Indian Institute of Technology Goa, Farmagudi, Goa -- 403 401, India.
}%
\author{Sujit Kumar Sahoo}
\email{sujit@iitgoa.ac.in}
\affiliation{ 
School of Electrical Sciences, Indian Institute of Technology Goa, Farmagudi, Goa -- 403 401, India.
}%
\author{Y. Sudhakar}%
 \email{sudhakar@iitgoa.ac.in}
\affiliation{ 
School of Mechanical Sciences, Indian Institute of Technology Goa, Farmagudi, Goa -- 403 401, India.
}%

\date{\today}

\begin{abstract}
Performing geometry-resolved simulations of flows over rough and porous walls is highly expensive due to their multiscale characteristics. Effective models that circumvent this difficulty are often used to investigate the interaction between the free-fluid and such complex walls. These models, by construction, employ an intrinsic averaging process and capture only macroscopic physical processes. However, physical experiments or direct simulations yield micro- and macroscale information, and isolating the macroscopic effect from them is crucial for rigorously validating the accuracy of effective models. Despite the increasing use of effective models, this aspect received the least attention in the literature. This paper presents an efficient averaging technique to extract macroscopic interface data from the flowfield obtained via direct simulations or physical experiments.  The proposed methodology employs a combination of signal processing and polynomial interpolation techniques to capture the macroscopic information. Results from the ensemble averaging are used as the reference to quantify the accuracy of the proposed method.  Compared to the ensemble averaging, the proposed method, while retaining accuracy, is cost-effective for rough and porous walls. To the best of our knowledge, this is the only averaging method that works for poroelastic walls, for which the ensemble averaging fails. Moreover, it applies equally to viscous- and inertia-dominated flows over irregular surfaces.

\end{abstract}

\maketitle

\section{Introduction}
\label{sec:intro}
Fluid flows over rough, porous and poroelastic walls are prevalent in nature~\cite{bottaro2019}. Understanding heat and fluid transport associated with such irregular walls is essential for developing biomimetic engineering devices. A well-known example is the development of skin-friction-reducing devices, called riblets, inspired by the denticles found on shark skins~\cite{garcia2011}. Optimal riblet geometries can yield skin-friction drag reduction of up to 10\% in turbulent flows~\cite{bechert1997}.

The characteristic feature of the aforementioned irregular surfaces is that the geometric microscale $l$ can be orders of magnitude smaller than the macroscale $H$ relevant for the surrounding fluid flow. This multiscale nature of the configuration, quantified by the scale separation parameter $\eta(=l/H)\ll 1$, makes geometry-resolved numerical simulations (called DNS from hereafter) practically impossible. This paper uses the terminology of a complex or irregular surface to denote a solid wall coated with rough, porous, or poroelastic features.

Due to the practical difficulty of performing DNS, effective models are widely employed to study the flow physics associated with rough/porous walls~\cite{rosti2018,khorasani2022}.  Beavers and Joseph~\cite{beavers1967}, based on their experiments on a channel flow over a porous block, proposed an empirical shear-dependent slip velocity boundary condition at the fluid-porous interface. This is one of the first effective models proposed. Due to the practical importance of flows over irregular walls, various improved models are rigorously derived based on  homogenisation\cite{jager1996boundary,bottaro2020effective, carraro2013,carraro2015effective,lacis2016,carraro2018effective,zampogna2020,eggenweiler2021effective,maruvsic2021,sudhakar2021}, volume-averaging \cite{ochoa1995,valdes2021novel,valdes2021}, 
or other frameworks~\cite{vafai1987,lacis2020,angot2017asymptotic,jain2022}.

Effective models are the only viable means of numerically studying fluid flows over irregular surfaces.  They describe the interaction between the irregular surface and the surrounding fluid flow via coupling conditions specified at the interface, which is not a physical one, but a mathematical interface introduced to describe the problem in the macroscopic sense. Hence, the interface is usually denoted as the \textit{nominal} interface.  The accuracy of an effective model is dictated by how closely these interface conditions capture the underlying transport phenomena across the interface.  A crucial step in developing an effective model is to test the accuracy of interface conditions by comparing the model predictions against results obtained from DNS or physical experiments.  A major challenge in carrying out such a validation study is as follows: the effective models capture only the macroscopic effects, while DNS/experiments provide data that contain microscopic fields in addition to macroscopic effects.  As highlighted in section~\ref{sec:problem}, this mandates that an appropriate averaging procedure be applied to the DNS/experimental data to isolate the macroscopic effects.  Despite the recent surge of interest in this subject, this aspect has received the least attention in the literature. This paper presents an efficient approach to address this issue. 

The averaging process has received the least attention in the literature because most existing studies considered a laminar~\cite{jager2001,goyeau2003, tachie2004,breugem2005,deng2005, agelinchaab2006,chandesris2006,zhang2009,liu2011,carraro2013,wu2018,lu2019,terzis2019,lu2020a,lu2020b,strohbeck2023} or turbulent~\cite{chandesris2009,kuwata2017,suga2018,chen2021,chu2021}  channel flow over a rough/porous medium.  In such a configuration, the macroscopic flow is unidirectional because the averaged velocity in the interface-normal direction is zero everywhere.  From the microscopic point of view, what happens in a single interface periodic unit is repeated for all units along the interface.  From the macroscopic point of view, the flow is homogeneous along the interface. The averaging procedure in such a configuration is straightforward: performing a \textit{single} DNS over a domain consisting of a \textit{single} roughness element and averaging the pressure/velocity along the flow direction is sufficient to obtain macroscopic fields in a laminar flow; for turbulent flows, the time-averaged fields can be averaged along the interface, due to the spatial homogeneity, to obtain macroscopic fields.  Eggenweiler and Rybak~\cite{eggenweiler2020} used volume averaging over a unit cell in the interior. At the interface, they applied same averaging but scaled the height of the averaging domain with 0.5 and 0.25. Rinehart et al.~\cite{rinehart2021} used a combination of plane and volume averaging to obtain macroscopic quantities. The transition from volume averaging used in the interior to the plane averaging at the interface was smooth. Using this new technique, they accurately extracted the macroscopic interface quantities and the the boundary layer within the porous domain. This is the only averaging method that has been demonstrated to work well with the interior boundary layer.

Extracting macroscopic data from DNS/experimental results for multidimensional flows over irregular surfaces is not straightforward.  The work of L$\bar{\text{a}}$cis \& Bagheri~\cite{lacis2016} is one of the first to compare DNS and a homogenized model for two-dimensional flows over a porous bed. Due to the non-availability of a reliable averaging process, they reported a comparison of the model against non-averaged DNS results. Another study on the flow over poroelastic media~\cite{lacis2017} also presented such comparisons. Due to the presence of microscopic oscillations in DNS,  rigorous quantification of the accuracy of the homogenized model was not possible from such comparisons.

A few studies have employed different procedures to obtain macroscopic fields from DNS.  Breugem \& Boersma~\cite{breugem2005} presented a comprehensive analysis of the turbulent flow over a permeable wall and compared the DNS results against those of an effective model. A volume averaging technique involving a weight function is used to obtain the macroscopic mean and turbulent fluctuating quantities.  Chandesris et al.~\cite{chandesris2013} used a similar averaging but adopted a weight function with increased regularity.
An investigation by Nair et al.~\cite{nair2018} focused on laminar flows over a porous flat plate employed averaging over a representative volume element~(REV).  The size of the REV equal to $20D$, where $D$ is the diameter of the circular solid inclusions used to represent the porous medium, was chosen to obtain macroscopic results from experiments and simulations.  Experimental studies on porous media~\cite{terzis2019,yang2019} used plane and volume averaging, respectively, to obtain macroscopic interface and interior quantities. A recent study of natural convection flow over a rough vertical wall used a moving averaging technique\cite{ahmed2022}. 

L$\bar{\text{a}}$cis et al.~\cite{lacis2020} utilized an ensemble averaging procedure to extract averaged fields from DNS results. This method differs from the volume averaging technique or plane averaging over the interface. Samples needed for the ensemble averaging are obtained by displacing the solid skeleton parallel to the interface by a uniform distance. This ensures that the generated samples cover the span of one microscopic length scale. This procedure's main feature is that it directly applies to multidimensional flows.  This averaging method is subsequently used in validating the effective models formulated in Cartesian~\cite{sudhakar2021} and Polar~\cite{jain2022} coordinates. Although it yielded smooth macroscopic fields even for multidimensional flows over irregular surfaces, generating samples for the ensemble averaging is daunting because each sample data requires a full DNS. This proves that the averaging is a computationally costly procedure. Our tests indicate that at least ten samples for each interface dimension are needed to eliminate the microscopic oscillations completely. While addressing three-dimensional flows over a complex surface, the computational cost becomes prohibitive because 100 expensive geometry-resolved simulations would be required. Inertia-dominated flows add additional complexities to this averaging procedure since they require longer time and more computational resources, owing to the underlying nonlinear nature of the governing PDEs.

In this paper, we present a computationally feasible procedure to extract macroscopic data at a nominal interface between an irregular surface and the surrounding two-dimensional free fluid flow.  We make use of the concepts of signal processing and polynomial interpolation to develop this method.
In contrast to the large number of samples required for ensemble averaging, the present method requires only two samples to provide accurate results. Thus the method is efficient when compared to the ensemble averaging. As shown in a later section, the proposed method works well for flows over poroelastic surfaces exhibiting large deformations, for which the ensemble averaging can not be applied. Moreover, we provide the source code and relevant data reported in this work in a public repository~\cite{bitbucket}.  For all the configurations reported in section~\ref{sec:results}, the microscopic interface data obtained using numerical simulations are included in the repository. The source code is written in MATLAB. It reads the microscopic data and computes the ensemble averaging as well as the macroscopic data using the present method.

This paper is organized as follows: In section~\ref{sec:problem}, we describe the appearance of microscopic oscillations in DNS,  and the importance of the averaging process for flows over complex walls. The methodology and the algorithm are detailed in \S~\ref{sec:method}. In section~\ref{sec:results}, for flows over rough, porous and poroelastic surfaces, results on extracting macroscopic data from DNS using the present method are presented; results obtained from our approach are compared to those from the ensemble averaging technique. Conclusions drawn from our study are summarised in \S~\ref{sec:conc}.
\section{Problem definition}
\label{sec:problem}
\begin{figure*}
\includegraphics[scale=1.0]{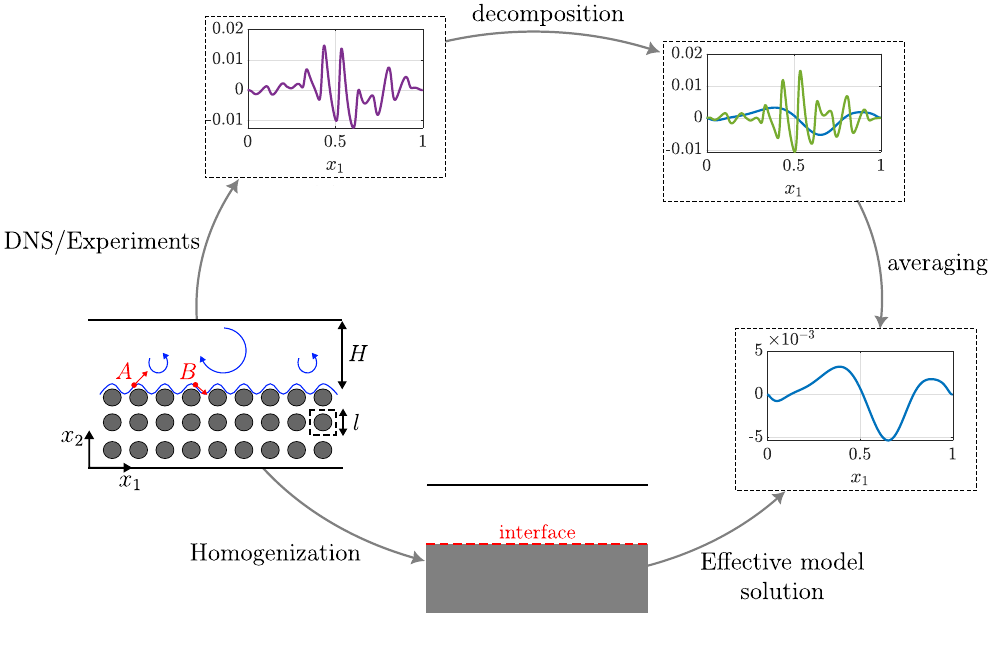}
\caption{Illustration of the averaging process. (a)~microscopic  description, (b)~velocity of a particle contains both microscale oscillations and macroscale variation, (c)~decomposition of data in (b) into macroscopic mean~(blue) and microscopic fluctuations~(green), (d)~averaging process yielding only macroscopic data, and (e)~macroscopic description by an effective model.}
\label{fig:probdef}
\end{figure*}

This section details the appearance of microscopic oscillations in DNS and the necessity of performing the averaging process to extract the macroscopic data. Let us consider a steady, laminar flow over an ordered porous medium, as shown in figure~\ref{fig:probdef}(a). The circular inclusions represent the solid portion of the porous domain. Here, the microscale $l$ denotes the length scale characterising the pore structure, and the characteristic length scale of the free-flow is $H$ (denoted in this paper as macroscale). The microscopic description can be obtained by performing DNS or conducting physical experiments. Such a description will yield both macroscopic as well as microscopic fields. This is explained as follows.

Let us assume that we are releasing a massless particle P in the domain at a point close to the interface between the free-fluid and the porous part. When it travels and approaches the fore surface of a circular inclusion, the particle experiences a positive vertical velocity (point A in figure~\ref{fig:probdef}a). The velocity component becomes negative when the particle moves past the aft surface of the inclusion (point B in figure~\ref{fig:probdef}a). The particle must have a zero vertical velocity between these two points. The vertical velocity's ``cyclic'' behaviour is repeated as it flows over each solid inclusion. This is the source of the appearance of the microscale oscillations, whose wavelength is always $l$. In addition, the particle itself travels within the macroscopic velocity field generated in the free fluid flow. Hence, a measurement of the vertical velocity of the particle in the microscopic description will contain both macroscopic and microscopic contributions, as shown in figure~\ref{fig:probdef}(b). Such an observation is valid for velocity components, pressure, and other variables in a fluid flow. Although the above illustration is presented for a porous wall, it also directly applies to rough and poroelastic surfaces.

Geometry-resolved simulations of flows over rough/porous walls are costly due to the large disparity of length scales quantified by the scale separation parameter. Volume-averaging or homogenization-based effective models, as shown in figure~\ref{fig:probdef}(e), are widely used to overcome this difficulty. Contrary to DNS, these models do not resolve the pore-scale geometry in the simulations. The fluid flow through the porous medium is represented by upscaled models~\cite{lage1998} like the Darcy, Darcy-Brinkmann, Forchheimer, or power law formulation. The interaction between the porous medium and the surrounding free-fluid region is modelled by specifying appropriate conditions at a nominal interface, as shown in figure~\ref{fig:probdef}(e). Effective models employ an intrinsic averaging procedure, and therefore by construction, they can capture only the macroscopic behaviour: the vertical velocity of the same particle mentioned above will appear as a smooth curve without microscopic variations, as shown in~figure~\ref{fig:probdef}(d).

The accuracy of any effective model is dictated by how accurately the respective interface conditions capture the transport phenomena across the nominal interface. Existing studies~\cite{lacis2020,sudhakar2021,jain2022} quantify the accuracy by comparing the interface quantities predicted by the model with those from DNS. Due to the widespread use of effective models, investigating their validity is crucial before using them in practical applications. In order to aid this validation, it is of utmost importance to devise a procedure to extract the macroscopic data (figure~\ref{fig:probdef}d) from the microscopic fields obtained via DNS/experiments (figure~\ref{fig:probdef}b).

The data obtained from DNS/experiments is of complex nature that they contain microscopic oscillations superimposed on the macroscopic field data. We can decompose a microscopic quantity~($Q$) into macroscopic~($\overline{Q}$) and fluctuations~($Q^\prime$), as shown below
\begin{equation}
Q(x,X) = \overline{Q}(X) + Q^\prime (x,X)\ \ \ \ \ \ \ \textrm{with\ \ \ \ \ }\overline{Q^\prime}=0,
\label{eqn:decomp}
\end{equation}
which is analogous to the Reynolds decomposition used in turbulent flows~\cite{davidson2015}.  However, contrary to turbulent fluctuations that are disordered and chaotic, the microscopic fluctuations appearing here are ordered in nature. Here, $Q$ can be any variable associated with the fluid flow. $x$ and $X$ represent fast and slow scales, respectively, in the context of multiscale homogenization~\cite{meibook}. $\overline{Q}$ contains the effect of microscale surface features on the length scale~($H$) of our interest, while details at the much smaller scales~($l$) are averaged out. Another point worthy of mentioning is the following. While the wavelength of the microscale field is $l$, its magnitude depends on the flow details. This explains why $Q^\prime$ is a function of both $x$ and $X$. Figure~\ref{fig:probdef}(c) illustrates the decomposition and underlines the above points.

The definition of the averaging mentioned in equation~\eqref{eqn:decomp} is delicate; it can dictate both the computational effort and the accuracy of obtaining macroscopic data. This work aims to devise an efficient method,  based on the knowledge from signal processing, to extract $\overline{Q}$ from $Q$. While ensemble averaging, the only procedure reported in the literature for multidimensional flows over irregular surfaces, requires many samples, the present method can yield accurate macroscopic data using only two samples.
\section{Methodology}
\label{sec:method}
In the previous section, we have introduced the decomposition of the microscopic quantity~($Q$) into macroscopic~($\overline{Q}$) and fluctuations~($Q^\prime$), as given by equation~\eqref{eqn:decomp}. In this work, we make use of additional observations to develop a simple averaging methodology and an associated algorithm to isolate the macroscopic effect from the data obtained via either DNS or experiments. These details are elaborated on in this section.

We can notice from the figure~\ref{fig:probdef} that $\overline{Q}$ is slowly varying over space.  However,  $Q$ carries high-frequency oscillations~($Q^\prime$) of varying amplitude riding on top of $\overline{Q}$. We observed in all our simulations, and as explained in the previous section,  $Q^\prime$ is periodic. Further, it can be expressed as the following multiplicative decomposition
\begin{equation}
Q^\prime (x,X)=w(X)t(x),
\end{equation}
where $w(X)$ is a smooth window that shapes the amplitude of these oscillating functions $t(x)$. Since $t(x)$ is a periodic function over space, we can decompose it as fundamental sinusoidal and its harmonics using the Fourier series.
\begin{equation}
t(x)=\sum_{k=1}^{\infty}\left[a_k\sin (k\omega_0x)+b_k\cos (k\omega_0x)\right],
\end{equation}
where $a_k$ and $b_k$ are the real Fourier series coefficients that set the amplitude of the respective oscillating sinusoidal. 

\begin{figure*}
\subfloat[]{\includegraphics[scale=0.3]{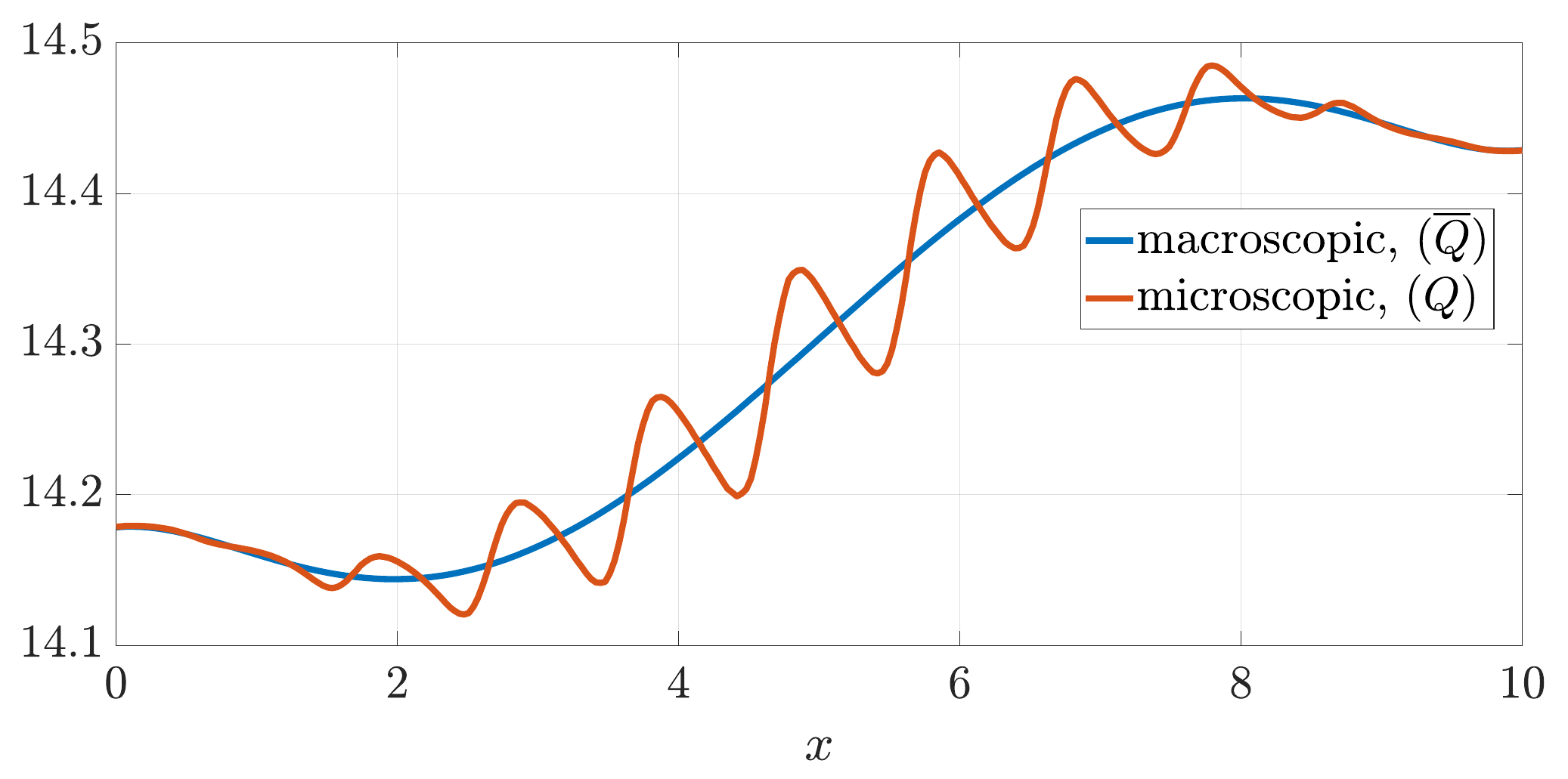}}\\
\subfloat[]{\includegraphics[scale=0.3]{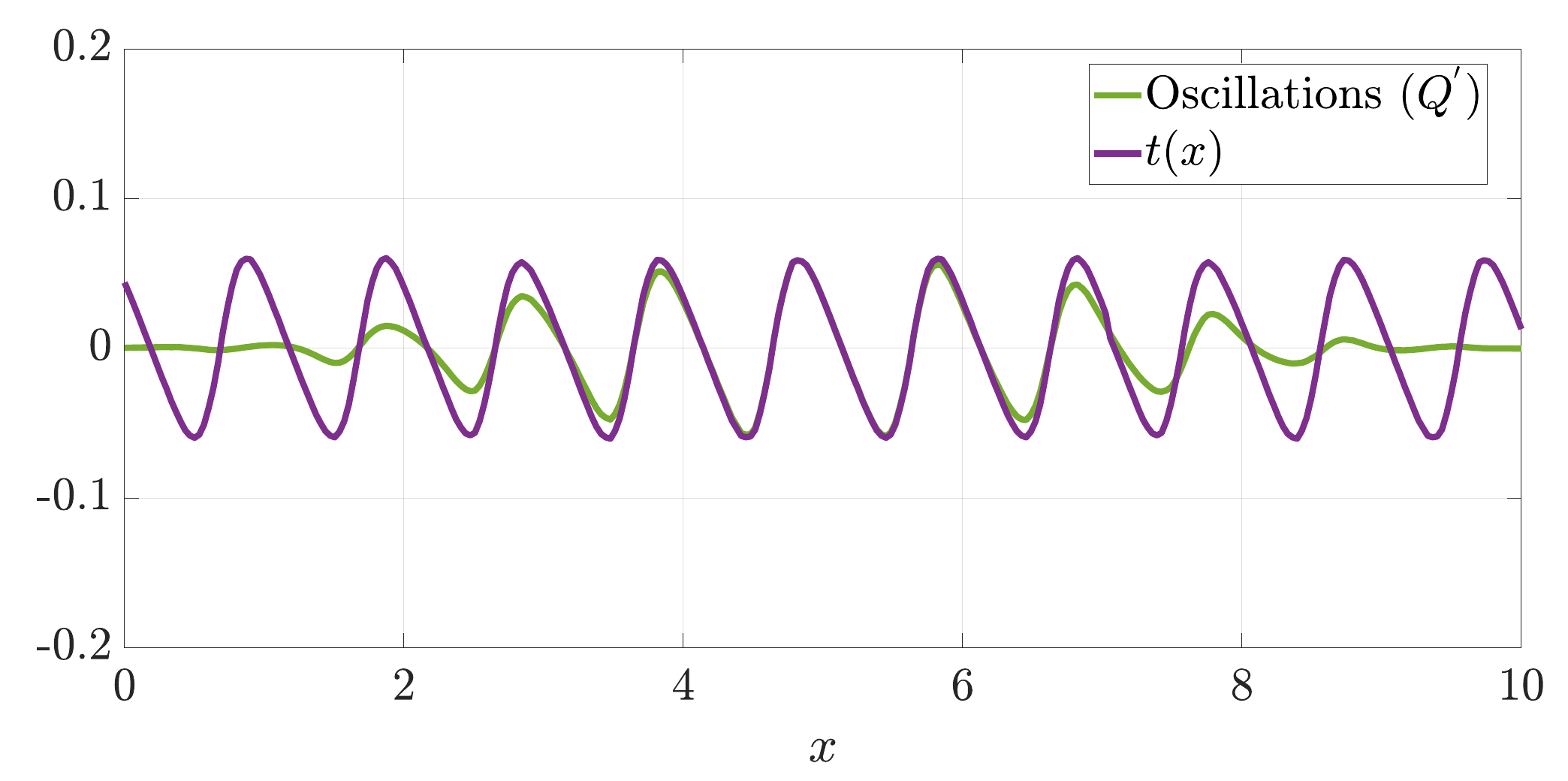}}\\
\subfloat[]{\includegraphics[scale=0.3]{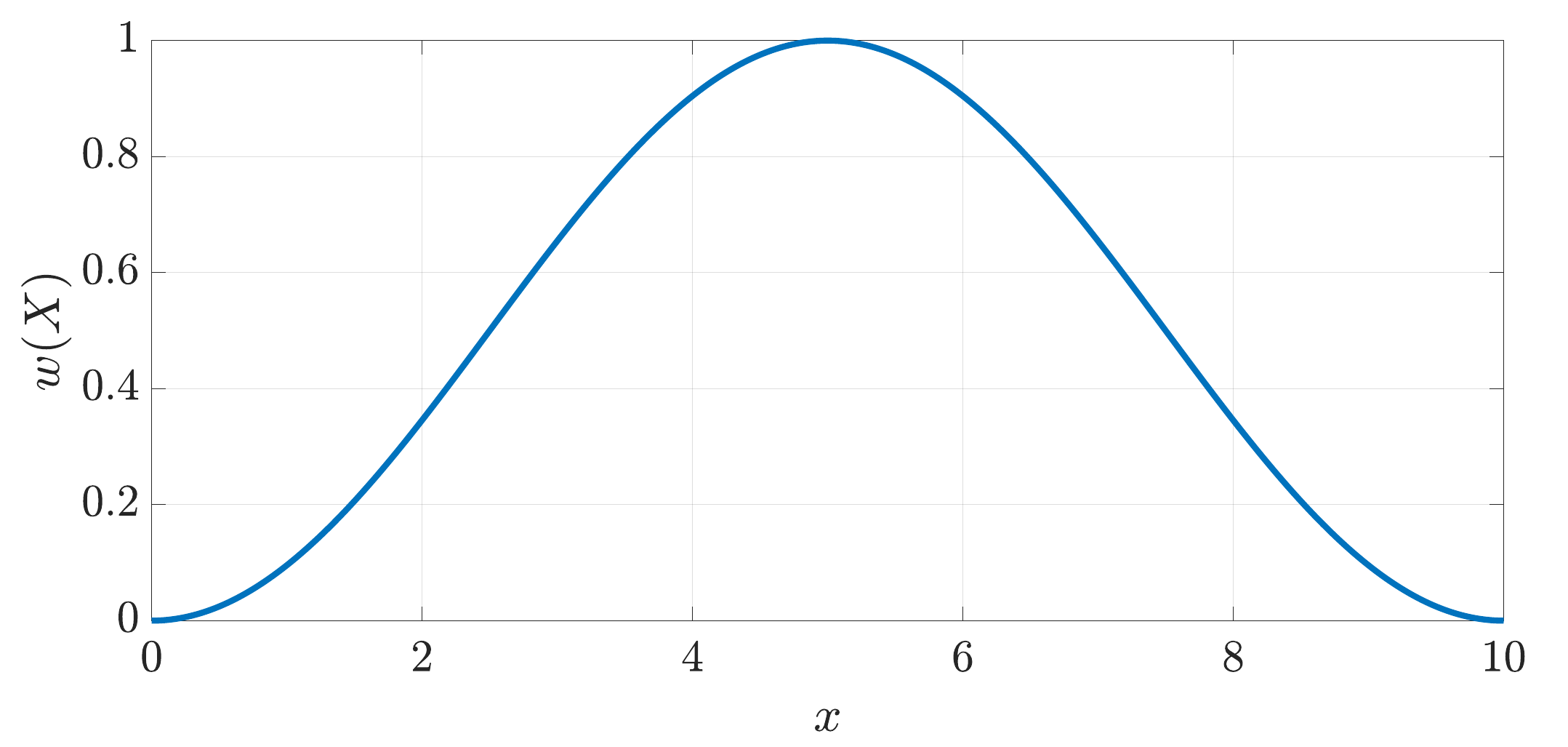}}
\caption{Illustration of the decompositions $Q(x,X)=\overline{Q}(X)+Q^\prime(x,X)$ and $Q^\prime(x,X)=w(X)t(x)$: (a)~macroscopic and microscopic quantities, (b)~fluctuations and its Fourier series representation with constant amplitude, and (c)~the window function.}
\label{fig:multdecomp}
\end{figure*}

To better understand the points mentioned above, we provide an example of the above decomposition. We consider the flow inside the lid-driven cavity with a porous bed, which is explained in section~\ref{sec:results}. The micro- and macroscales for this problem are $l=1$ and $H=10$. Figure~\ref{fig:multdecomp}(a) illustrates the distribution of pressure along the interface in the microscopic description, and its macroscopic version obtained using an ensemble averaging technique\cite{lacis2020,sudhakar2021,jain2022}. Additional details regarding ensemble averaging have been discussed in the previous section~\ref{sec:intro}. It is directly evident from figure~\ref{fig:multdecomp} that $Q^\prime$ exhibits a periodic waveform with the fundamental oscillation frequency $\omega_0=1/l$, where $l$ is the microscale. The multiplicative decomposition of $Q^\prime$ into a periodic function with constant amplitude $t(x)$, and the window function $w(X)$ that shapes the amplitude can be understood from figures~\ref{fig:multdecomp}(b) and (c).

\begin{figure}
\subfloat[]{\includegraphics[scale=0.3]{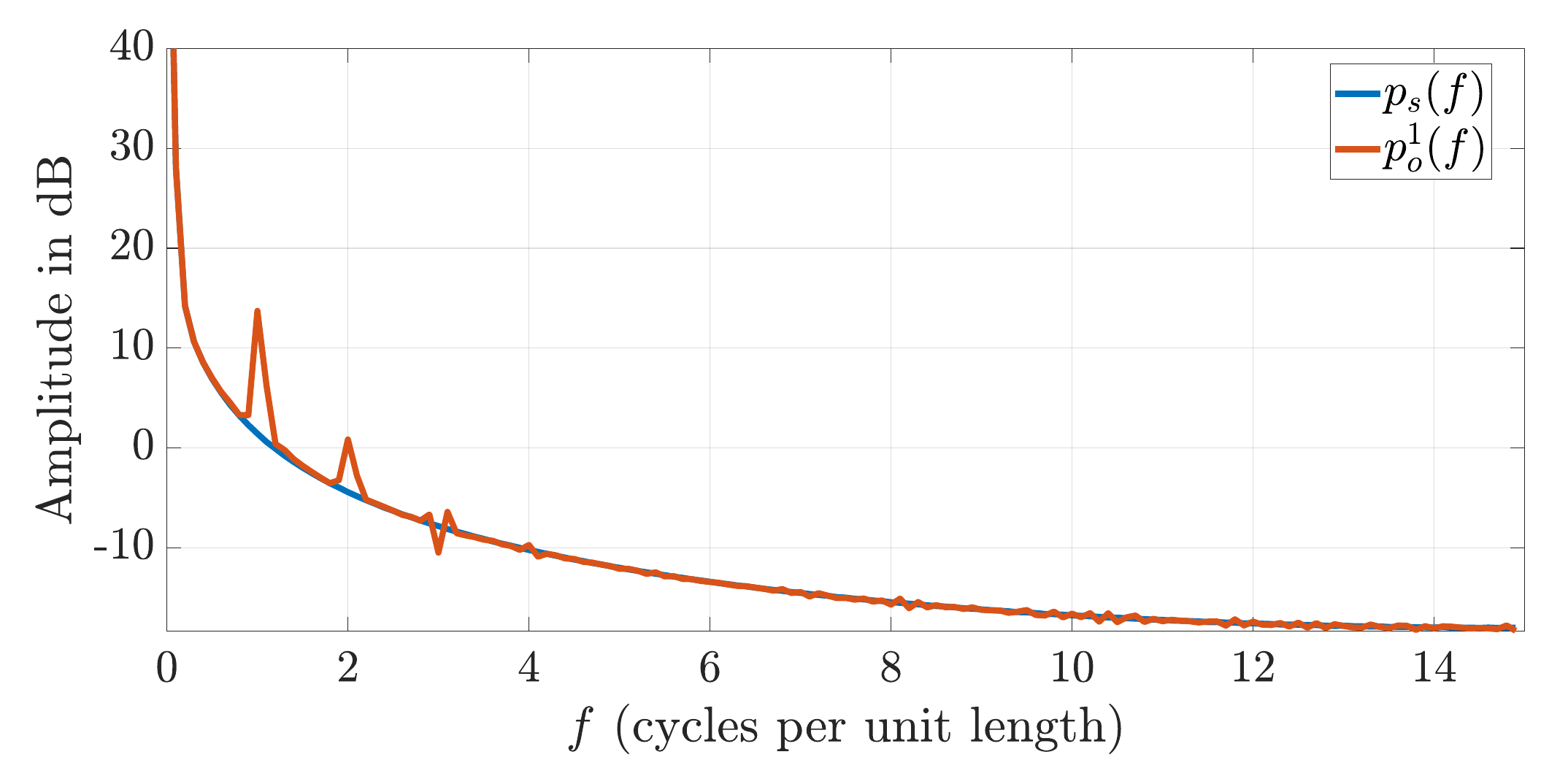}}\\
\subfloat[]{\includegraphics[scale=0.3]{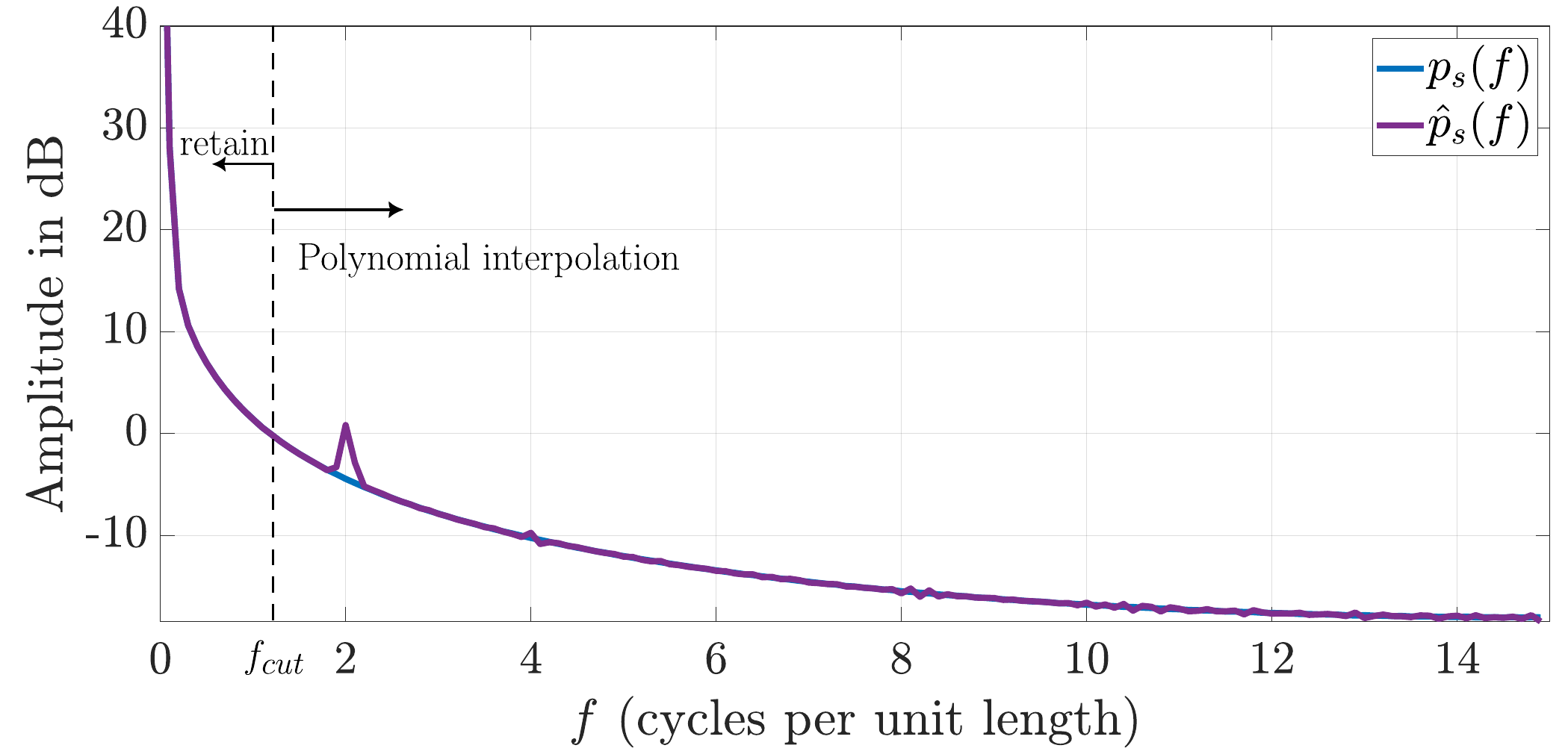}}
\caption{Frequency spectrum of (a)~DNS~($p_0^1$), ensemble averaged~($p_s$) data, and (b)~$p_s$ and two-sample averaged~($\hat{p}_s$) data along the interface.  Amplitudes are in dB scale, which is $20\log_{10}$. A brief overview of the method is also given in~(b). We introduce a cut-off frequency~($f_{cut}$) between $\omega_0$ and 2$\omega_0$.  The segment below $f\le f_{cut}$ is retained in the frequency spectrum, while the portion $f> f_{cut}$ is smoothed using polynomial interpolation.}
\label{fig:freqspec}
\end{figure}

\begin{figure}
\includegraphics[scale=0.45]{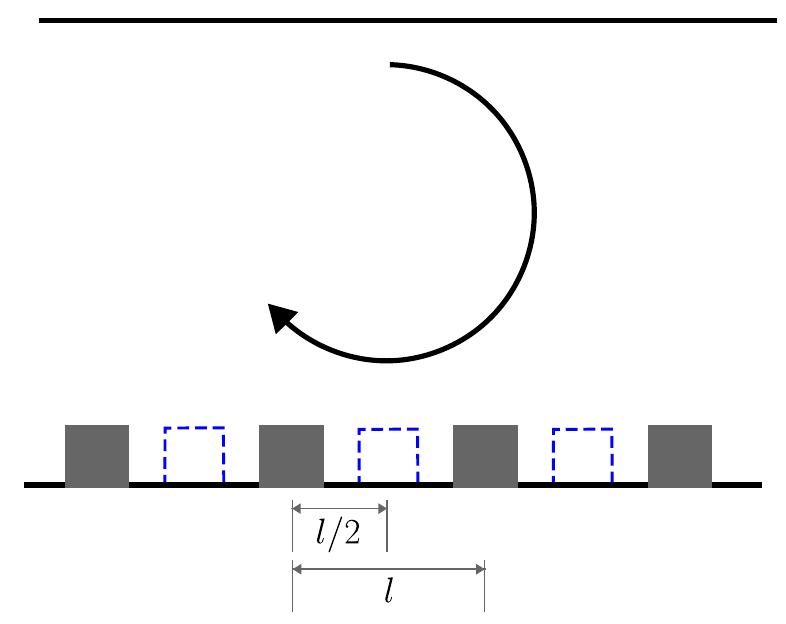}
\caption{Illustration of the two samples used in the present method. Grey filled rough elements represent the geometry for the first sample, and they are moved by a distance $l/2$ (blue dotted) for the second sample.}
\label{fig:twoconfig}
\end{figure}

An improved understanding of the nature of microscopic and macroscopic data will help us to design an efficient averaging method. In order for this, we investigate the frequency spectrum of relevant quantities. We can notice from figure~\ref{fig:freqspec}(a) that while the macroscopic field has a smooth spectrum,  the microscopic data contains undesirable spikes in the spectrum due to the presence of oscillations $Q^\prime$. With careful observation, we can conclude that these spikes occur at the harmonic frequencies of the oscillations: $\omega_0$, 2$\omega_0$, 3$\omega_0$,  4$\omega_0$ and so on. ensemble averaging, performed using data collected by moving the microscopic inclusions to a uniform distance between sampling points, nullifies these unwanted spikes through destructive superposition. However, performing ensemble averaging is highly expensive. The present work aims to achieve the same effect of such averaging with only two samples. To apply our method, DNS of two configurations, between which the microscale features are moved by a distance of $l/2$, are performed to get the necessary two samples. These two configurations are depicted in figure~\ref{fig:twoconfig} for the flow over a rough wall.

The smoothing in the present method takes place in three steps:
\begin{enumerate}
\item We obtain the average interface values from the two samples described in figure~\ref{fig:twoconfig}. This averaging cancels out the most dominant first harmonic (and other visible odd-harmonics), as can be seen from the spectrum plotted in figure~\ref{fig:freqspec}(b).
\item The remaining harmonics in the frequency spectrum are removed through a polynomial interpolation.
\item The macroscopic signal in the physical space is obtained by performing an inverse Fourier transform of the smoothed spectrum.
\end{enumerate}
As will be explained in section~\ref{sec:results}, the present method produces results as accurate as that of the expensive ensemble averaging technique.

Algorithm 1 presents the implementation details of the above steps. Let $p_0^1(x)$ and $p_0^2(x)$ indicate microscopic interface data obtained from the two samples, shown in figure~\ref{fig:twoconfig}. Our objective is to get the averaged macroscopic representation ${p}_s(X)$. Input parameters $f_{cut}$ and  $deg$ are the frequency above which the polynomial interpolation is applied and the degree of the polynomial, respectively. We take the mean of the two microscopic signals as the initial guess for ${p}_s(X)$ (line~2 of the algorithm), and perform Fourier transform to get its spectrum~(line~3). As already discussed using figure~\ref{fig:freqspec}(b), the most dominant first harmonic is removed in the mean signal. The present method removes the remaining undesirable spikes from the frequency spectrum as follows; the piece of (smooth) spectrum with frequency $f<f_{cut}$ is retained, and the remaining part of the spectrum is reconstructed by removing spikes using polynomial interpolation. This step is explained in lines 4 and 5 of the algorithm and illustrated in figure~\ref{fig:freqspec}(b). Finally, the desired macroscopic signal in the physical space is obtained by performing an inverse Fourier transform~(line~9).


\begin{algorithm*}
\DontPrintSemicolon
	\caption{Extraction of macroscopic data from two microscopic signals} 
   \quad \textbf{Inputs}: $p_0^1(x)$, $p_0^2(x)$,  $f_{cut}$, and $deg$\;
	  \quad \textbf{Initialization}: $\hat{p}_s(x)=[p_0^1(x)+p_0^2(x)]/2$\;
	  \quad \textbf{Compute Fourier transform}: $\hat{P}_s(f)=\mathcal{F}[\hat{p}_s(x)]$\;
	   \quad \textbf{Perform smoothing by the splitting $\hat{P}_s(f)$ into two parts}:\;
\quad\quad \quad\quad$
P_s(f) = \begin{cases} 
\hat{P}_s(f) &\mbox{for } 0\le f \le f_{cut}  \\
P_s(f)= \arg \min \sum\limits_{f>f_\text{cut}}{\left[\hat{P}_s(f) - P_s(f)\right]^2} \text{ s.t. } P_s(f) = \sum_{k =0}^{deg}{a_k f^k} & \mbox{for } f>f_{cut}
\end{cases}
$\;
	 \quad\textbf{Output}: $p_s(X) = \mathcal{F}^{-1}[P_s(f)]$
\end{algorithm*}

In all our test cases, we found that setting the polynomial degree~($deg$) between 5 to 20 gave accurate results. In this work, we used $deg=10$ and $f_{cut}=1.5\omega_0$.
\section{Results}
\label{sec:results}
Effective models approximate a complex surface as a rough, porous or poroelastic wall. In this section, we apply the proposed method to all these cases to demonstrate its broader applicability. Two test cases are widely used to examine the accuracy of effective models: channel flow and lid-driven cavity.  We consider the latter because the wall-normal velocity is zero in a channel flow, and based on our experience, averaging this velocity component is challenging. Results from the ensemble averaging are used as the reference to quantify the accuracy of the proposed method.

Flows over porous and rough walls are used to show that the method proposed in this paper produces the same results as that of the ensemble averaging technique but at a much lower computational cost. It will be shown that the ensemble averaging process encounters inherent difficulties while dealing with poroelastic surfaces exhibiting large deformation. The method proposed here works well even for such cases.
\subsection{Flow over a porous medium}
\begin{figure}
\includegraphics[scale=0.52]{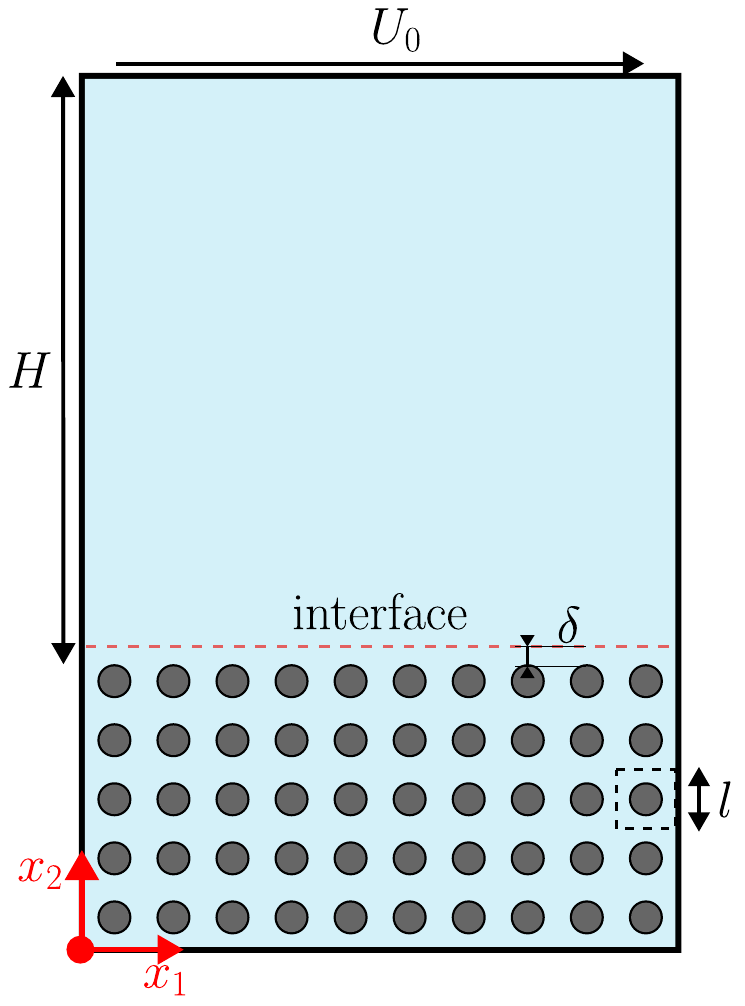}
\caption{Simulation setup for the flow over a porous medium. }
\label{fig:porgeom}
\end{figure}

As the first test case, we consider the flow over a porous medium. The simulation setup, which involves a lid-driven cavity with a porous bed, is depicted in figure~\ref{fig:porgeom}.  The circular inclusions represent the solid portion of the porous medium, and we consider Stokes flow over this isotropic ordered porous wall.  The porosity is 0.75, $l$=1~m, $H$=10~m,  and $U_0=10$~m/s. We choose the interface height, $\delta=0.1l$, to report the results.

\begin{figure*}
\subfloat[]{\includegraphics[scale=0.55]{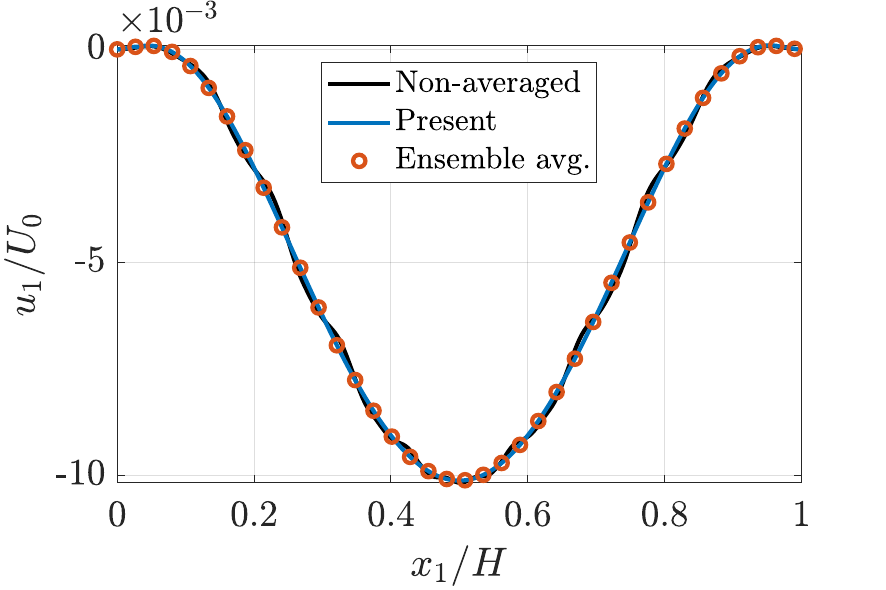}} 
\subfloat[]{\includegraphics[scale=0.55]{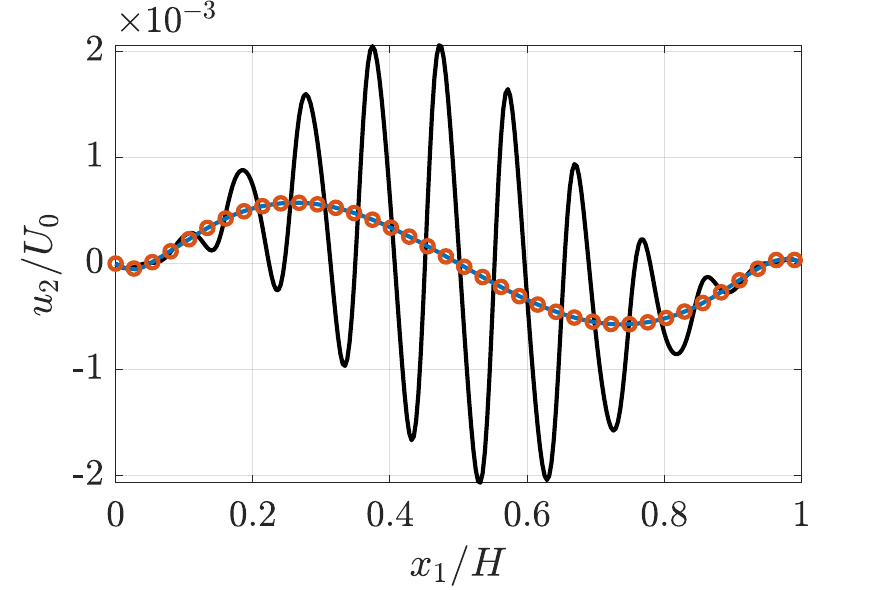}}
\caption{Averaged data obtained along the interface for the porous test case using ensemble averaging and the present method (a)~tangential velocity (b)~wall-normal velocity.}
\label{fig:porousres}
\end{figure*}

We aim to obtain the macroscopic variation of velocity components along the interface. Since this is the first example, we present the non-averaged microscale tangential and wall-normal velocities along the interface in figure~\ref{fig:porousres}. We see that the presence of the solid inclusions introduces variations in the field variables over macro- and micro-length scales. While the high-frequency variation represents microscale phenomena, the low-frequency component contains macroscopic information. Effective models are constructed to compute macroscopic quantities, and as stated earlier, this work aims to extract such quantities. For this problem, the ensemble averaging technique requires approximately ten samples to completely eliminate low-frequency oscillations. Hence, ten geometry-resolved simulations need to be performed to employ such averaging. Plots of averaged velocity variation obtained using the ensemble averaging and the proposed method, shown in figure~\ref{fig:porousres}, indicate that the proposed method produces results as accurate as the ensemble averaging. It is to be noted that the present method can extract the macroscale variation only by using two samples, and thus it reduces the computational cost significantly.

In order to quantify the accuracy of the present method, when compared to the ensemble averaging, we compare the minimum tangential velocity and the maximum wall-normal velocity at the interface. These quantities were used to describe the accuracy of effective models in earlier studies~\cite{lacis2020,sudhakar2021,jain2022}. These values, reported in Table~\ref{tab:porous}, clearly show that the present method gives the same values as the ensemble averaging, with only a negligible difference. Thus, we can conclude that the technique reported in this paper produces as accurate results as that of the expensive ensemble averaging.

\begin{table*}
\caption{\label{tab:porous} Comparison of the minimum tangential velocity and the maximum wall-normal velocity, for the lid-driven cavity with the porous bed,  obtained using the present method and the ensemble averaging.}
\begin{ruledtabular}
\begin{tabular}{cccc}
\multicolumn{2}{c}{minimum $(u_1/U_0)$} & \multicolumn{2}{c}{maximum $(u_2/U_0)$} \\\cline{1-2} \cline{3-4}
ensemble avg. & Present & ensemble avg. & Present \\
$-1.011888\times 10^{-2}$ & $-1.011953\times 10^{-2}$ & $5.7381\times 10^{-4}$ & $5.7335\times 10^{-4}$ \\
\end{tabular}
\end{ruledtabular}
\end{table*}

Although we presented only the results for an isotropic porous medium here,  we verified that the averaging process described in this paper also works for anisotropic porous inclusions.

\begin{figure*}
\subfloat[]{\includegraphics[scale=0.55]{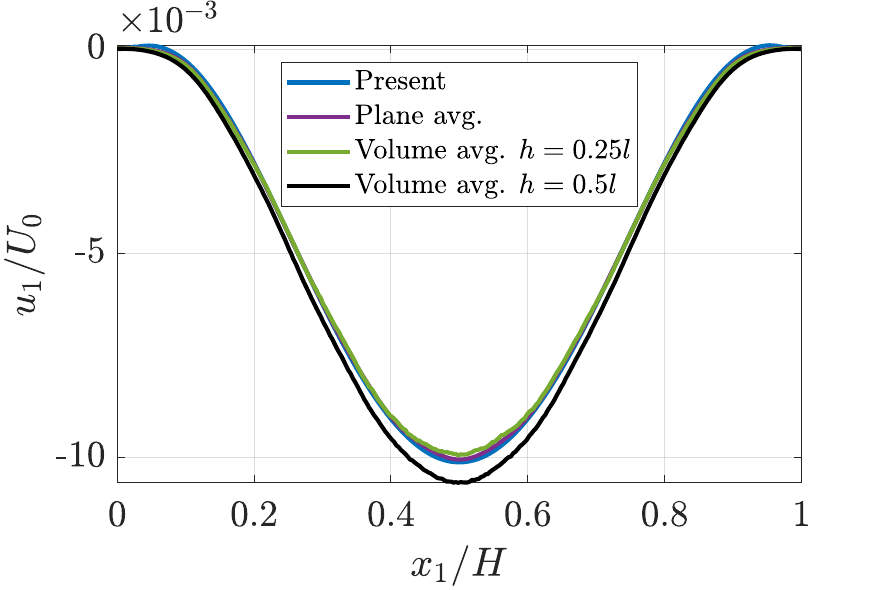}}
\subfloat[]{\includegraphics[scale=0.55]{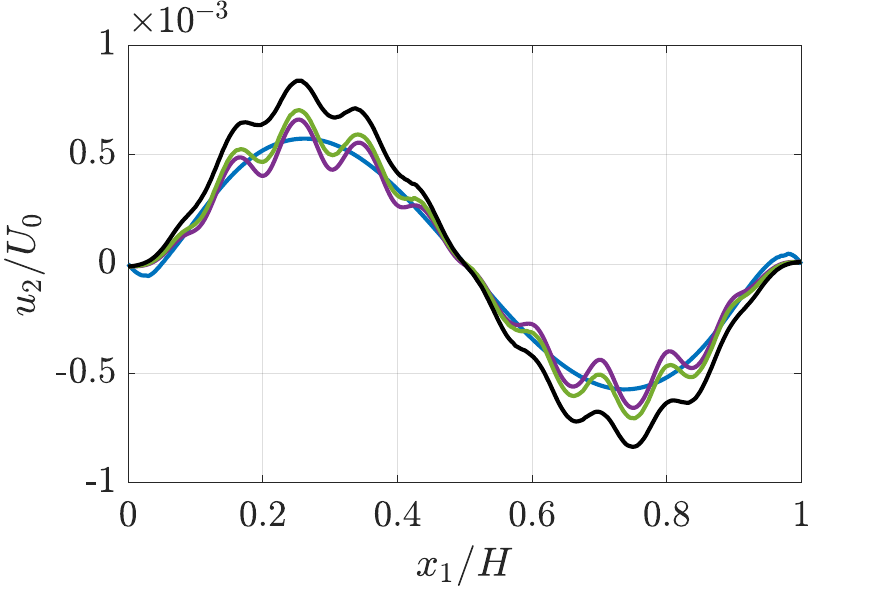}}
\caption{Comparison of the present method, plane averaging and volume averaging for the flow over the porous medium: (a)~tangential velocity (b)~wall-normal velocity. Volume averaging is performed using two different heights~($h$) of the averaging domain.}
\label{fig:volavg}
\end{figure*}

In the introduction, we stated that the averaging procedure to extract macroscopic data is straightforward for a unidirectional channel flow with a porous bed or a rough wall. It is of interest to compare the procedure used in the literature for channel flow with that of the present method. We chose the following two averaging procedures for comparison
\begin{align}\label{eqn:newavg1}
\textrm{Plane averaging:} \quad &\overline{Q}(\tilde{x},y_i)=\frac{1}{l}\bigintsss_{\tilde{x}-l/2}^{\tilde{x}+l/2}{Q(x)dx_1},\\
\textrm{Volume averaging:}\quad &\overline{Q}(\tilde{x},y_i)=\frac{1}{hl}\bigintsss_{y_i-h/2}^{y_i+h/2}\bigintsss_{\tilde{x}-l/2}^{\tilde{x}+l/2}{Q(x)dx_1 dx_2}.\label{eqn:newavg2}
\end{align}
In the above equations, the averaging is done at the point $(\tilde{x},y_i)$. Here, we denoted the location of the interface as $x_2=y_i$.  The plane averaging given in equation~\eqref{eqn:newavg1} is performed over a microscopic length scale $l$.  The volume averaging described above represents the averaging over a unit cell when $h=l$.  Eggenweiler and Rybak~\cite{eggenweiler2020} used $h=0.25l$ and $h=0.5l$ at the interface.

We compare the accuracy of the plane and the volume averaging with the present method, for the Stokes flow over the porous domain described above.  Macroscopic interface data obtained from all these three methods are presented in figure~\ref{fig:volavg}. Since the ensemble averaging and the present method produced same results, we omitted the former from figures for clarity. The plane averaging very closely matches with the tangential velocity obtained from the present method, as shown in figure~\ref{fig:volavg}(a). Macroscopic data captured by performing volume averaging approaches the reference values when the height of the averaging domain~($h$) is reduced. This is expected because in theory, shrinking $h=0$ is equivalent to performing plane averaging. In contrast to the tangential velocity, neither plane nor volume averaging technique is able to completely eliminate the microscopic oscillations in the transpiration velocity, as shown in figure~\ref{fig:volavg}(b). This could be because the magnitude of the microscopic oscillations of $u_2$ is as large as the macroscopic values. The plane and the volume averaging uses information only from a single DNS, which may not be sufficient to annihilate the oscillations. The present method, as described earlier, utilize interface data from two geometry-resolved simulations, and is able to accurately extract the macroscopic data. 

The aforementioned discussion provides confirmation to our earlier statement that extracting macroscopic transpiration velocity is more challenging than the tangential velocity. 
\subsection{Flow over rough walls}
\begin{figure}
\includegraphics[scale=0.6]{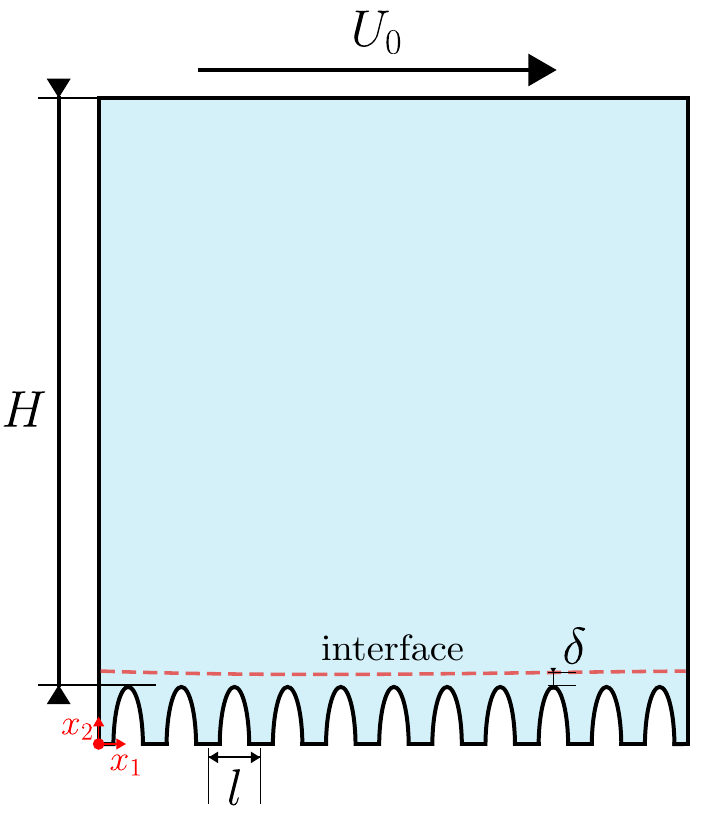}
\caption{Simulation setup for the flow over a rough wall.}
\label{fig:roughgeom}
\end{figure}

Having demonstrated the applicability of the present method to flow over a porous medium, in this section,  we consider the flow within a lid-driven cavity in which the bottom wall is covered with ordered roughness elements, as shown in figure~\ref{fig:roughgeom}. In the previous section, we investigated only Stokes flow. This section aims to verify the applicability of the proposed method to flows in which the inertial effects are non-negligible. Within this rough-wall cavity, we perform simulations at a Reynolds number ($Re=U_0H/\nu$) of 0~(Stokes flow), 100 and 1000. Semi-major and semi-minor axes of the roughness elements are 0.25$l$ and $l$, respectively. The scale separation parameter, $\eta=0.1$. We choose the interface height, $\delta=0.1l$, to report the results.

\begin{figure*}
\subfloat[]{\includegraphics[scale=0.55]{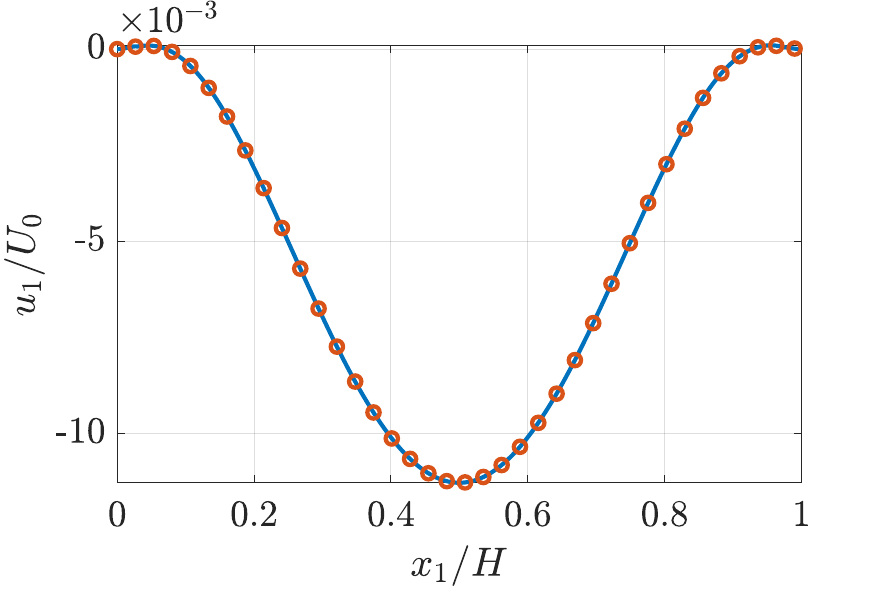}} 
\subfloat[]{\includegraphics[scale=0.55]{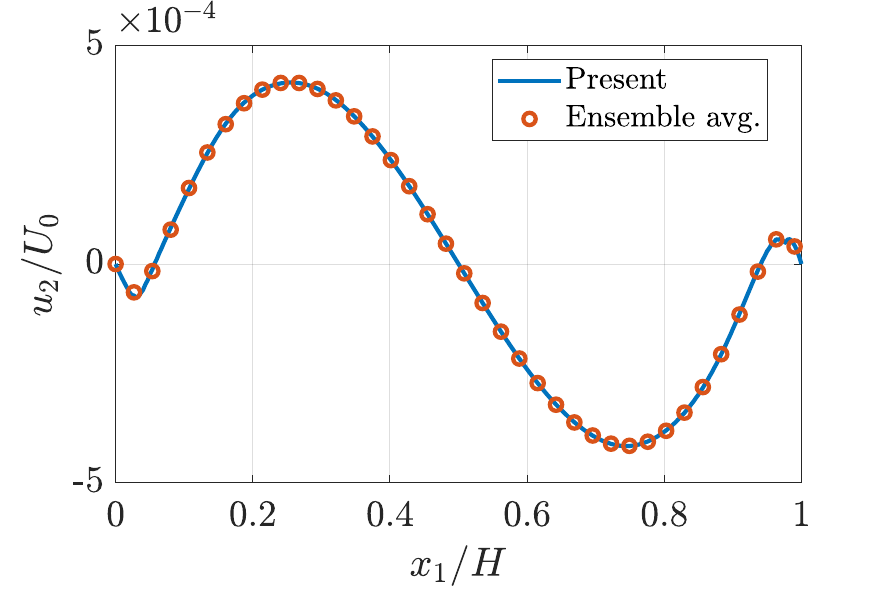}}\\
\subfloat[]{\includegraphics[scale=0.55]{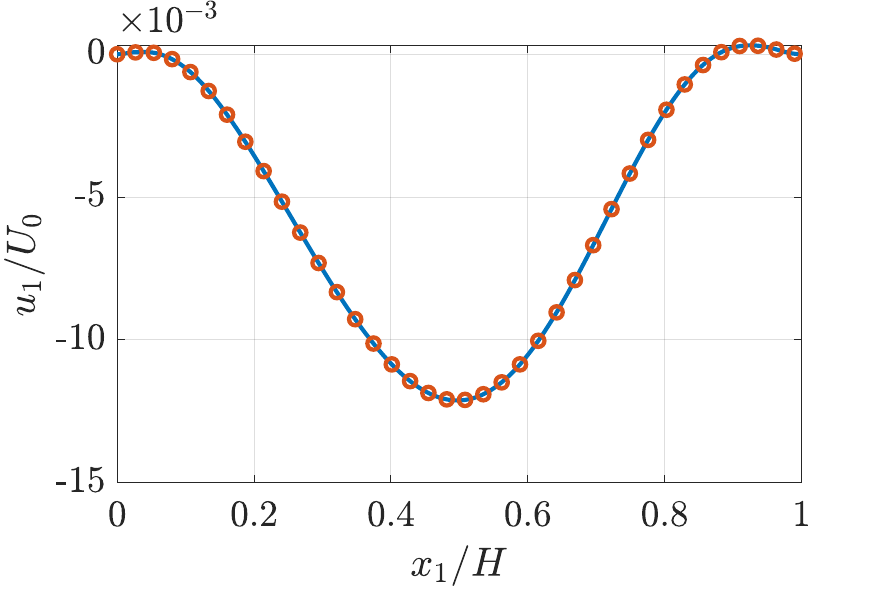}} 
\subfloat[]{\includegraphics[scale=0.55]{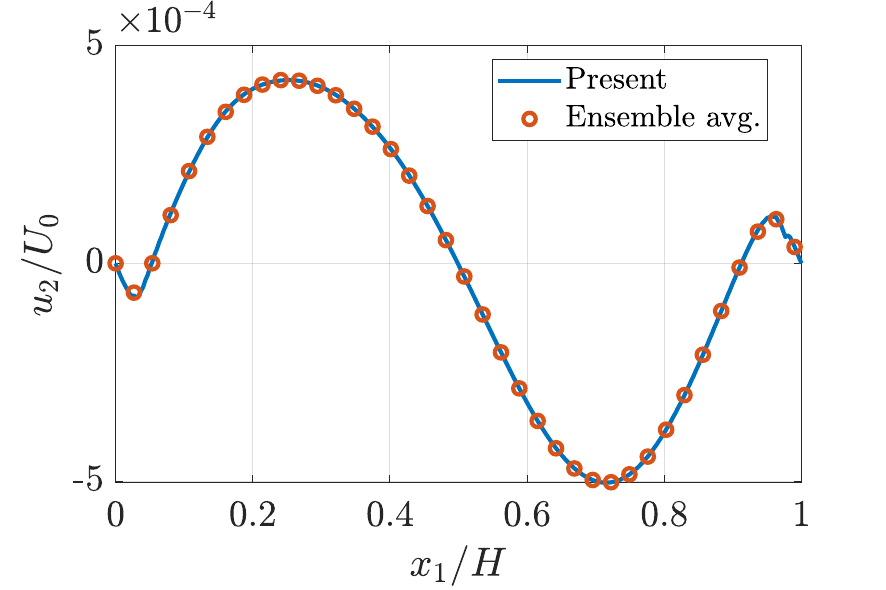}}\\
\subfloat[]{\includegraphics[scale=0.55]{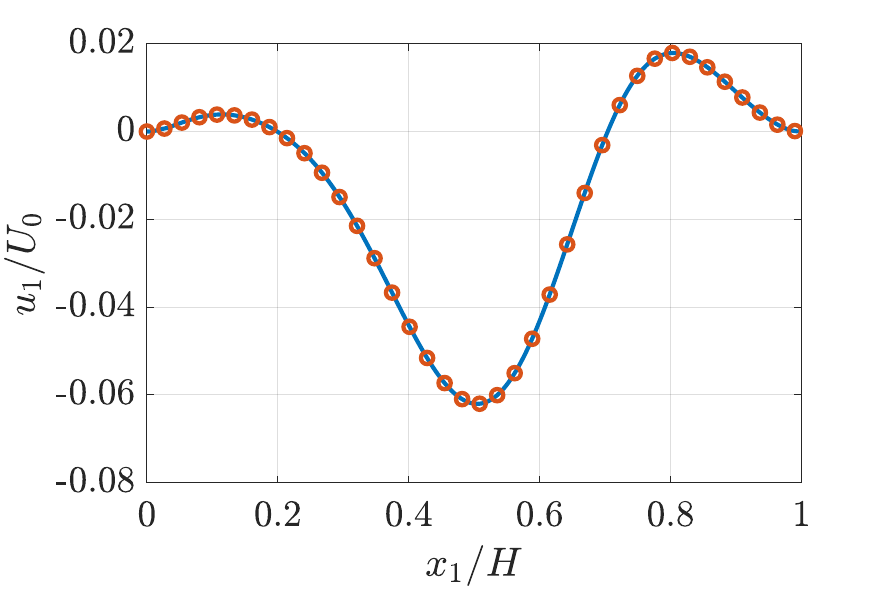}} 
\subfloat[]{\includegraphics[scale=0.55]{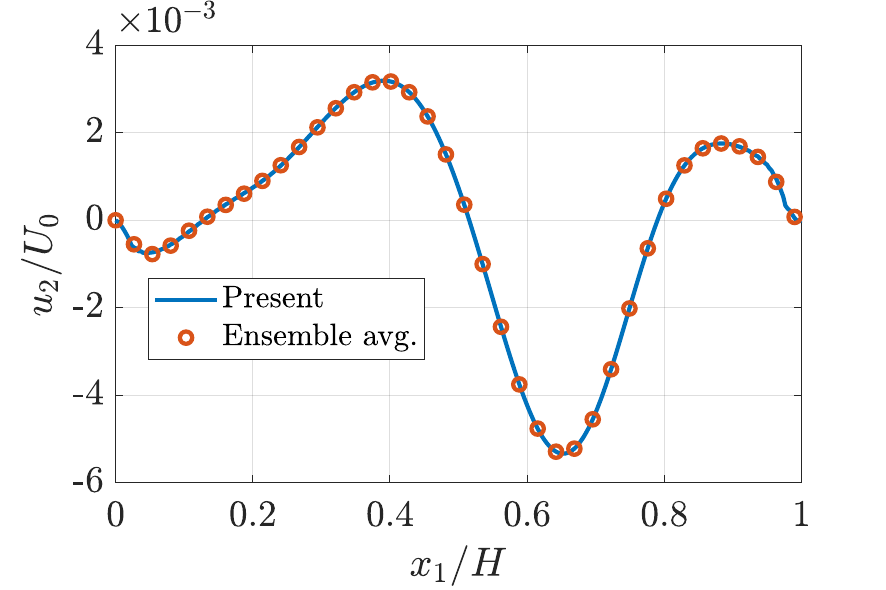}}\\
\caption{Averaged data obtained along the interface for the cavity with rough wall using ensemble averaging and the present method: tangential velocity~(left) and wall-normal velocity~(right). Top row contains results of Stokes flows, middle and bottom rows are $Re=100$ and 1000 results, respectively.}
\label{fig:roughres}
\end{figure*}

\begin{table*}
\caption{\label{tab:rough} Comparison of the minimum tangential velocity and the maximum wall-normal velocity, for the lid-driven cavity with rough wall, obtained using the present method and the ensemble averaging.}
\begin{ruledtabular}
\begin{tabular}{ccccc}
$Re$ & \multicolumn{2}{c}{minimum $(u_1/U_0)$} & \multicolumn{2}{c}{maximum $(u_2/U_0)$} \\\cline{2-3} \cline{4-5}
& ensemble avg. & Present & ensemble avg. & Present \\
0        & $1.127388\times 10^{-2}$ & $1.127491\times 10^{-2}$  & $4.155904\times 10^{-4}$    &  $4.162649\times 10^{-4}$ \\
100   & $1.211736\times 10^{-2}$ & $1.211848\times 10^{-2}$  & $5.028988\times 10^{-4}$   &  $5.032383\times 10^{-4}$\\
1000 & $6.202675\times 10^{-2}$ & $6.202663\times 10^{-2}$  & $5.337153\times 10^{-3}$ & $5.334945\times 10^{-3}$\\
\end{tabular}
\end{ruledtabular}
\end{table*}

The DNS produced velocity and pressure fields that contain macro- and micro-scale variations, similar to the flow over the porous medium reported in the previous section.  Since we discussed this point already, we present only the averaged results in this section. 

\begin{figure*}
\subfloat[]{\includegraphics[width=0.4\linewidth]{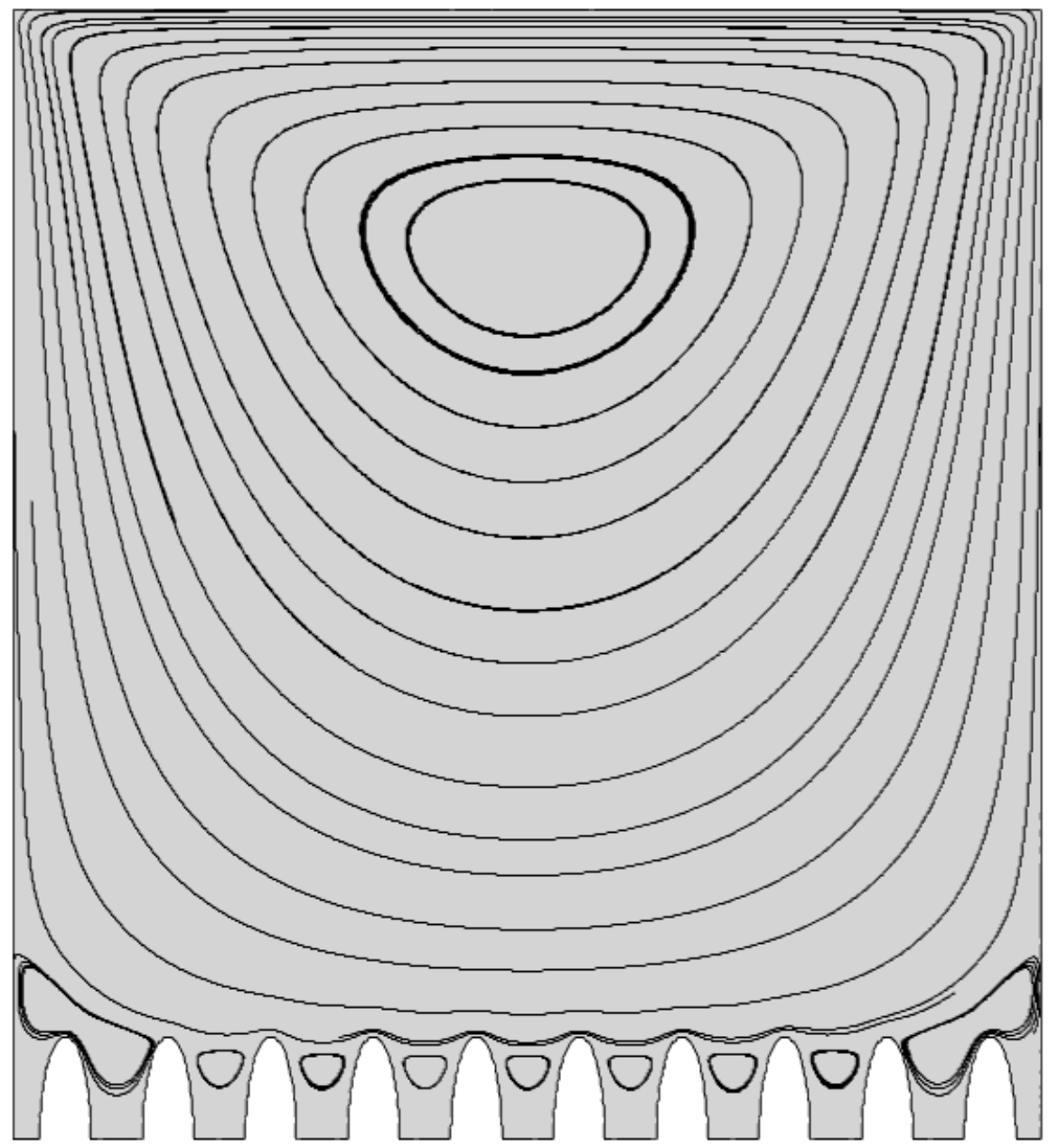}} \hspace{0.5cm}
\subfloat[]{\includegraphics[width=0.4\linewidth]{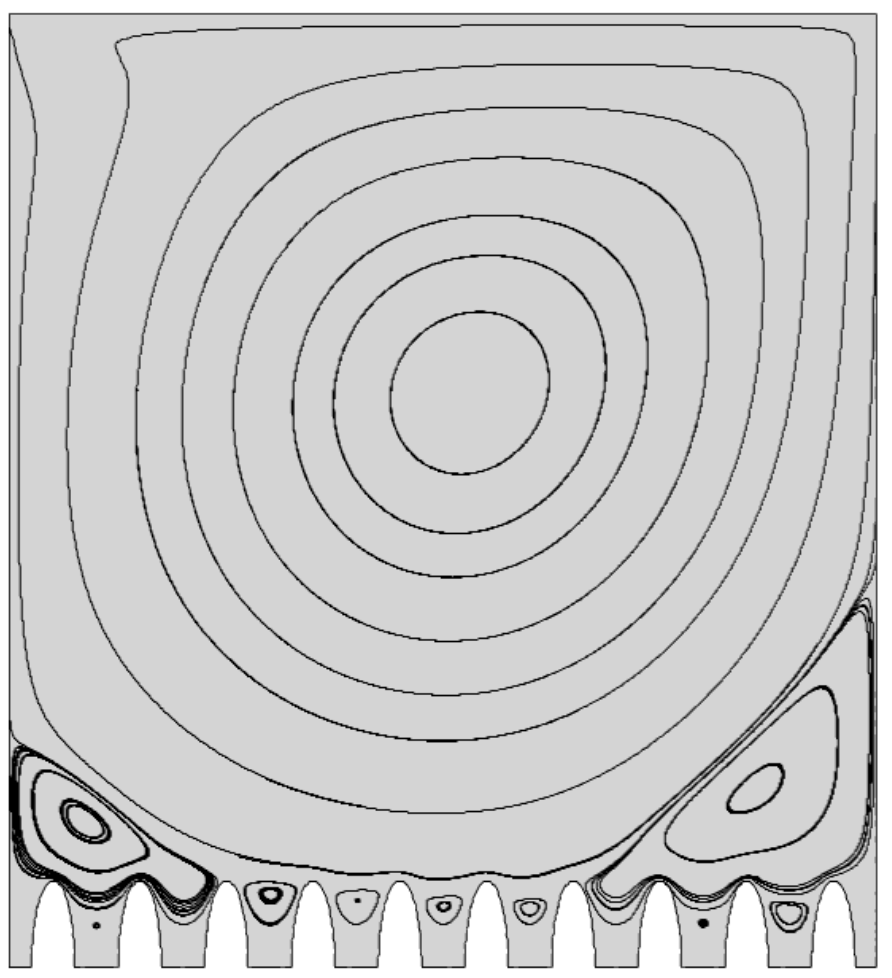}}\\
\caption{Streamlines in the cavity with rough bottom (a)~Stokes flow (b)~$Re=1000$.}
\label{fig:roughstreamline}
\end{figure*}

The tangential and wall-normal interface velocities for the Stokes flow, $Re=100$ and 1000 are presented in figure~\ref{fig:roughres}. As can be seen from the figure, for all simulations, the proposed method produces results as accurate as the ensemble averaging. Additional confirmation of the accuracy is provided by comparing the minimum tangential and the maximum wall-normal interface velocity. These values presented in table~\ref{tab:rough} clearly show that the difference between the present and the ensemble averaging method is negligible. From these results, we can conclude that the method presented in this paper works well for both viscous-dominated and inertia-dominated laminar flows over rough surfaces.

Similar to the porous wall, the ensemble averaging required approximately ten samples to fully eliminate the microscale oscillations. The present method, by construction, only requires two samples. The reduction in computational time is more significant for $Re=1000$ than for the Stokes flow.  This is because the Picard iteration or the Newton-Raphson method requires more iterations to converge\cite{engelman1981} for a large $Re$, while the governing equations are linear for the Stokes flow. 

Before concluding this section,  we relate the velocity profiles at the interface to the flowfield occurring within the cavity. The non-averaged streamline plots for the Stokes flow and at $Re=1000$ are presented in figure~\ref{fig:roughstreamline}. Due to linearity, the Stokes flow is symmetric with respect to the vertical line passing through the cavity centre. As a result, the velocity profiles are also symmetric, as shown in figure~\ref{fig:roughres}(a) and (b). Moreover, the main vortex formed within the cavity induces negative shear along the whole interface for Stokes flow. Hence, the tangential velocity is also negative, as explained by the well-known slip-velocity model~\cite{beavers1967}. In contrast, the flowfield is unsymmetric for $Re=1000$, hence the velocity profiles~(figure~\ref{fig:roughres}e and f). One of the significant differences at $Re=1000$ is the appearance of corner vortices, which induce a positive shear along a portion of the interface, and corresponding positive tangential velocity regions at either end of the interface. A larger $u_1$ at the right end indicates that the respective corner vortex is stronger. In recent works~\cite{lacis2020,sudhakar2021}, considering Stokes flow in the interface region, the following model for wall-normal velocity is proposed
\begin{equation}
u_2\propto-\frac{\partial u_1}{\partial x_1}.
\end{equation}
As can be seen from figure~\ref{fig:roughres}(e) and (f), $u_2$ is positive when $\frac{\partial u_1}{\partial x_1}<0$, and vice versa along the entire interface.  This points out that although the above model is obtained theoretically by eliminating inertial effects,  for the considered example, this relation is valid even when the interface Reynolds number, $Re_s=u_1^\textrm{max}l/\nu>6$.
\subsection{Flow over a poroelastic wall}
Results presented in the previous section for porous and rough walls indicate that the approach proposed in this paper can efficiently extract macroscopic data from DNS. Ensemble averaging, although expensive, works well for these examples. This section presents a test case for which the ensemble averaging fails, but the present method works without any issues.

\begin{figure*}
\includegraphics[scale=0.8]{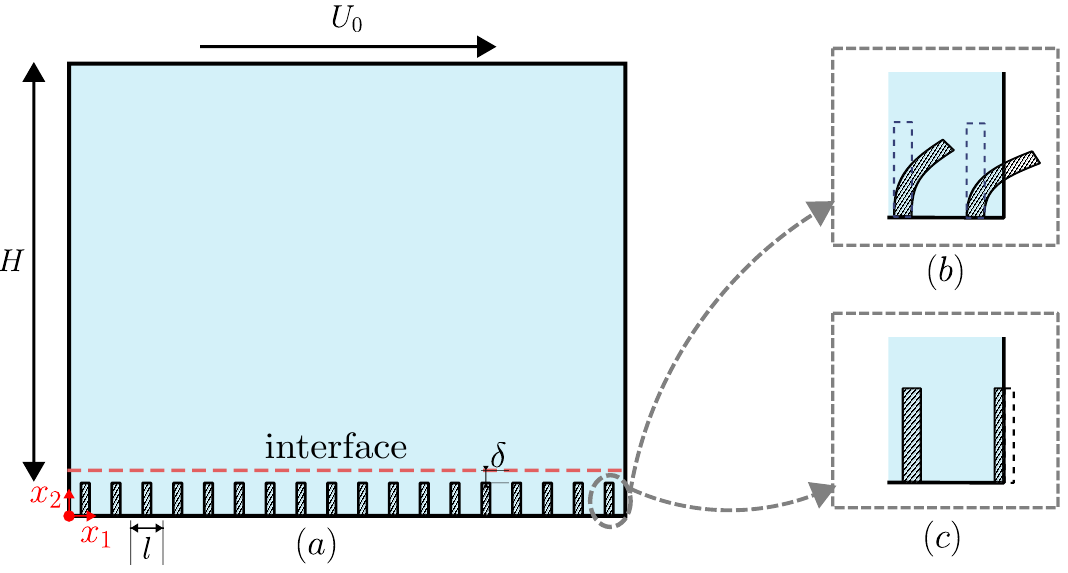}
\caption{Flow over a poroelastic wall.  (a)~Geometry of the domain and filaments, Possible failure of ensemble averaging due to (b)~filaments moving out of the domain due to the elastic deformation, (c)~part of the filament on the boundary. }
\label{fig:hairygeom}
\end{figure*}

The configuration considered is presented in figure~\ref{fig:hairygeom}(a). The lid-driven cavity problem is considered in which the bottom wall is coated with rod-like hairy filaments. The filaments are elastic, and hence they deform due to the pressure and viscous forces acting on them. We consider Stokes flow within the cavity. The parameters considered are $l=1$~m, $H=10$~m,  dimension of the hairy filament $0.5\times 0.1$~m$^2$; the ratio of solid to fluid density is 10. Neo-Hookean material model is used for the solid, and the interface height $\delta=0.1l$.  These parameters lead to a large deformation of the filaments, and should be modelled using a two-way coupled fluid-structure interaction technique.  We simulated this problem with the COMSOL Multiphysics\textsuperscript{\tiny\textregistered} (version 6.1) software using a partitioned coupling approach. We performed a quasi-steady FSI simulation, implying that the elastic filament attains an equilibrium position due to the fluid loads exerted. 

As discussed earlier,  moving the microscale geometry by one microlength scale is a prerequisite to applying the ensemble averaging.
This requirement is challenging to satisfy, for the poroelastic problem, due to the following two factors: (i)~during sampling, the filament can reach close to the cavity walls, and due to the subsequent fluid-induced deformation,  the filament may touch or even move out of the domain boundary as shown in figure~\ref{fig:hairygeom}(b), and (ii)~while moving the filament by equal distances between samples, only a part of the filament may lie on one wall and the remaining filament will be associated with the opposite cavity wall as illustrated in figure~\ref{fig:hairygeom}(c). The latter case is not an exception and happens for rough and porous walls discussed in earlier sections. Due to the elastic deformation of the filament, these scenarios are challenging to handle. Both these challenges stem from the large deformation of filaments, and they limit the applicability of the ensemble averaging only to small deformation cases.

\begin{figure*}
\subfloat[]{\includegraphics[scale=0.55]{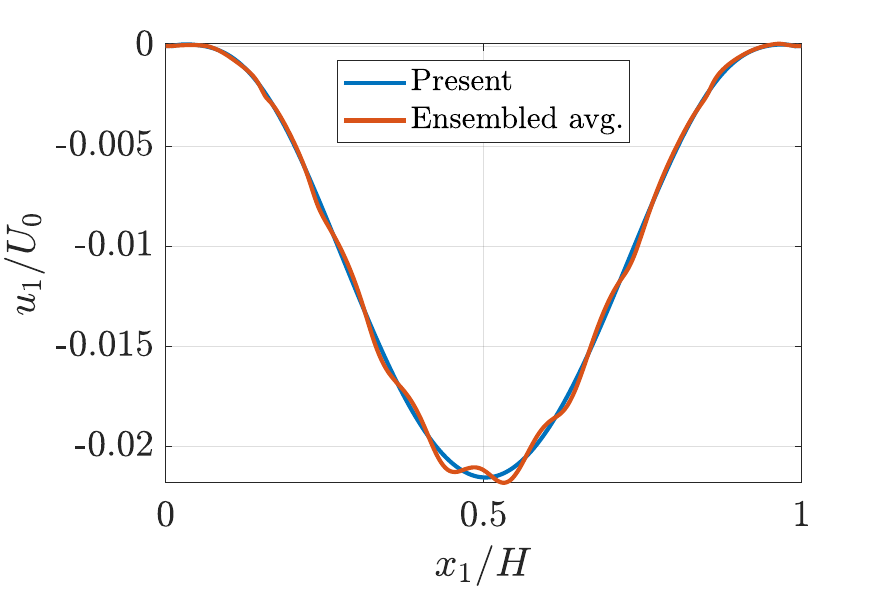}} 
\subfloat[]{\includegraphics[scale=0.55]{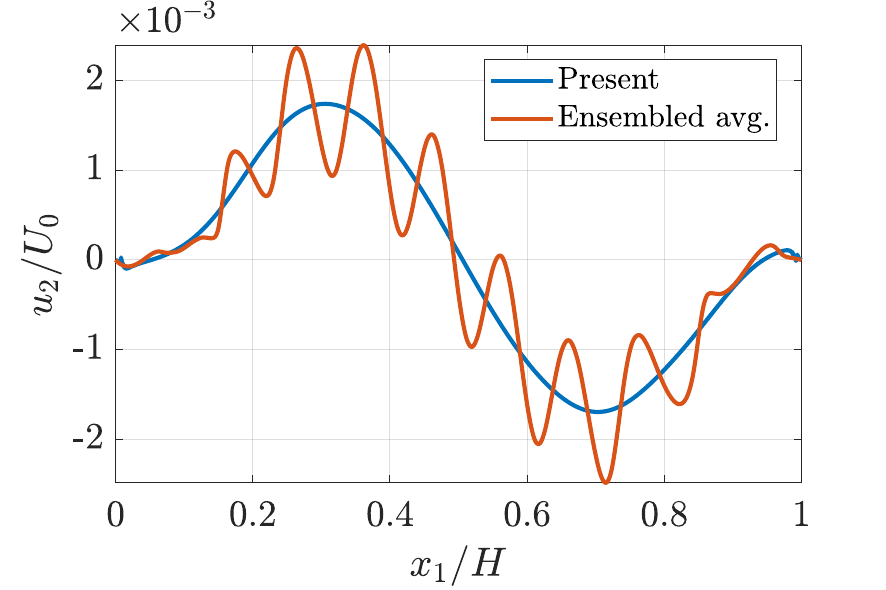}}
\caption{Averaged data obtained along the interface for the poroelastic test case using ensemble averaging and the present method (a)~tangential velocity (b)~wall-normal velocity.}
\label{fig:flexres}
\end{figure*}

As mentioned earlier, the configuration considered leads to large deformations of the filaments.  Due to the abovementioned challenges associated with the ensemble averaging, 102 samples are generated that cover 51\% of the microscopic length scale. The averaged tangential and wall-normal velocity along the interface are presented in figure~\ref{fig:flexres}. Since the complete microscopic length scale is not covered, the averaging does not extract the macroscopic fields from DNS results, and the microscopic oscillations prevail even after performing averaging. As shown in figure~\ref{fig:flexres}, while the oscillations in the tangential velocity are negligible, the wall-normal velocity exhibits significant microscale oscillations. 

The proposed method, with only two samples, is able to recover the smooth macroscopic data as shown in figure~\ref{fig:flexres}. This is also true for wall-normal velocity that exhibited large oscillations with ensemble averaging. Even if ensemble averaging would have worked, it would have required performing at least ten DNS and in this particular case, ten multiphysics geometry-resolved simulations. Performing such simulations is computationally very expensive. This highlights the potential of the proposed method in extracting macroscopic data from geometry-resolved simulations, even in cases for which ensemble averaging fails.

\begin{figure*}
\includegraphics[scale=0.4]{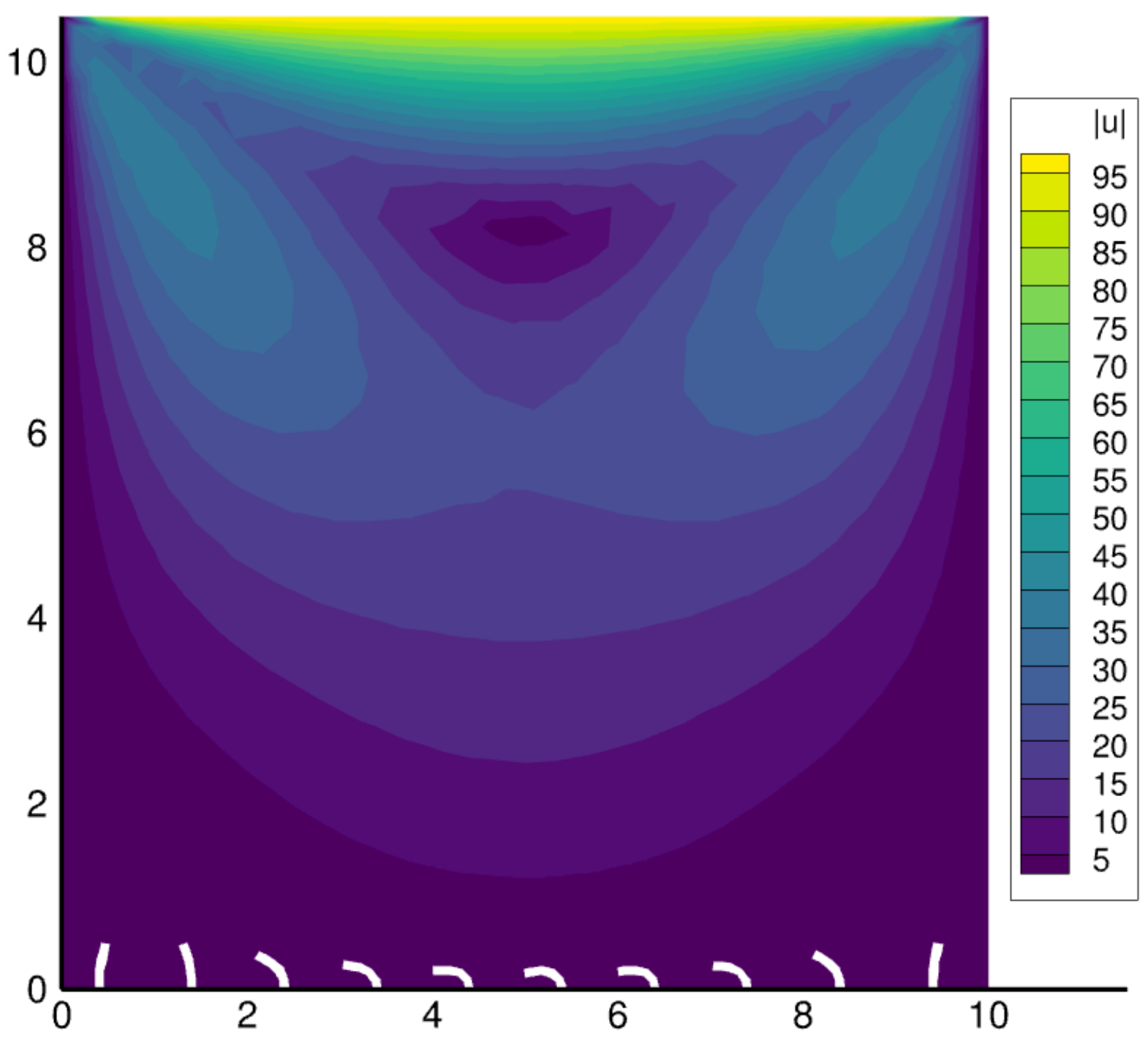}
\caption{Contours of velocity magnitude and elastic deformation of the hairy filaments for the flow over a poroelastic medium.}
\label{fig:hairycont}
\end{figure*}

\begin{figure*}
\subfloat[]{\includegraphics[scale=0.55]{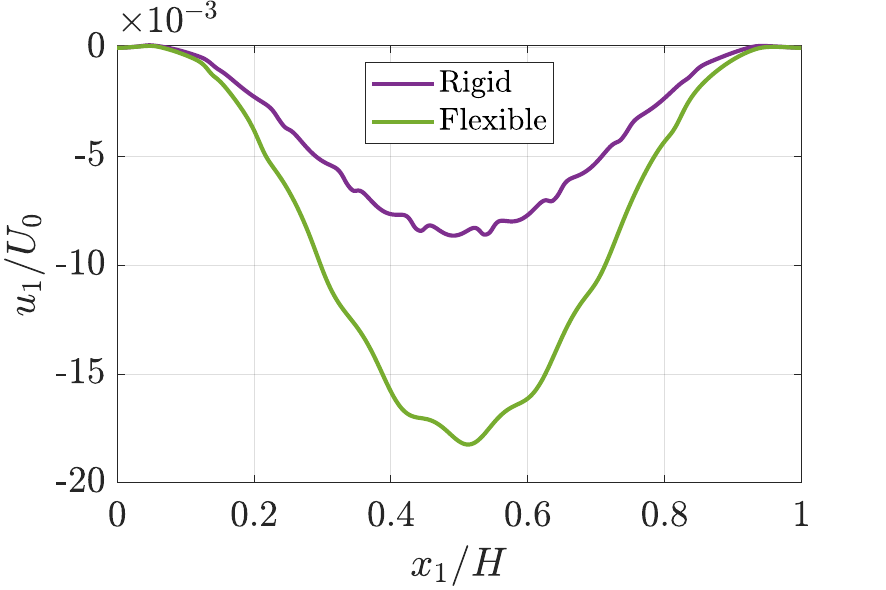}} 
\subfloat[]{\includegraphics[scale=0.55]{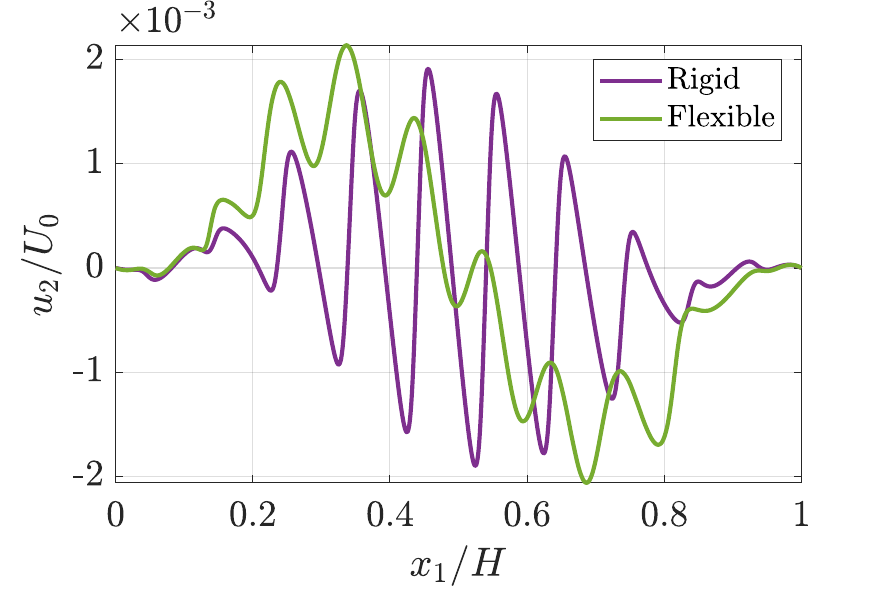}}
\caption{Comparison of interface velocities for rigid and flexible hairy filaments (a)~tangential velocity, (b)~wall-normal velocity.}
\label{fig:rigidflex}
\end{figure*}

The results presented above lead to the conclusion that the proposed method can also apply to non-ordered rough/porous surfaces. To explain this, let us consider the results depicting the deformed configuration of the filaments, as shown in figure~\ref{fig:hairycont}. Parameters that characterise the macroscopic description of this problem are the slip length~($\mathcal{L}$) for the tangential velocity and the transpiration length~($\mathcal{M}$) for the wall-normal velocity~\cite{lacis2020}. If the filaments were rigid, both $\mathcal{L}$ and $\mathcal{M}$ would be constant along the entire interface length for viscous-dominated flows considered here. However, in case of the flexible filaments, the local fluid flow dictates the deformed configuration of each of them. The amount of elastic bending is different for each filament, and hence $\mathcal{L}$ and $\mathcal{M}$ are non-constant functions along the interface. This is illustrated in figure~\ref{fig:rigidflex}, which shows the velocity components along the interface for flexible and rigid filaments. The difference between these two indicates the spatial variation of $\mathcal{L}$ and $\mathcal{M}$, even in the absence of inertial effects. Since such variation is a characteristic feature of non-ordered rough/porous surfaces, we envisage that our method works equally well for non-ordered surfaces. 

\begin{figure*}
\includegraphics[scale=0.4]{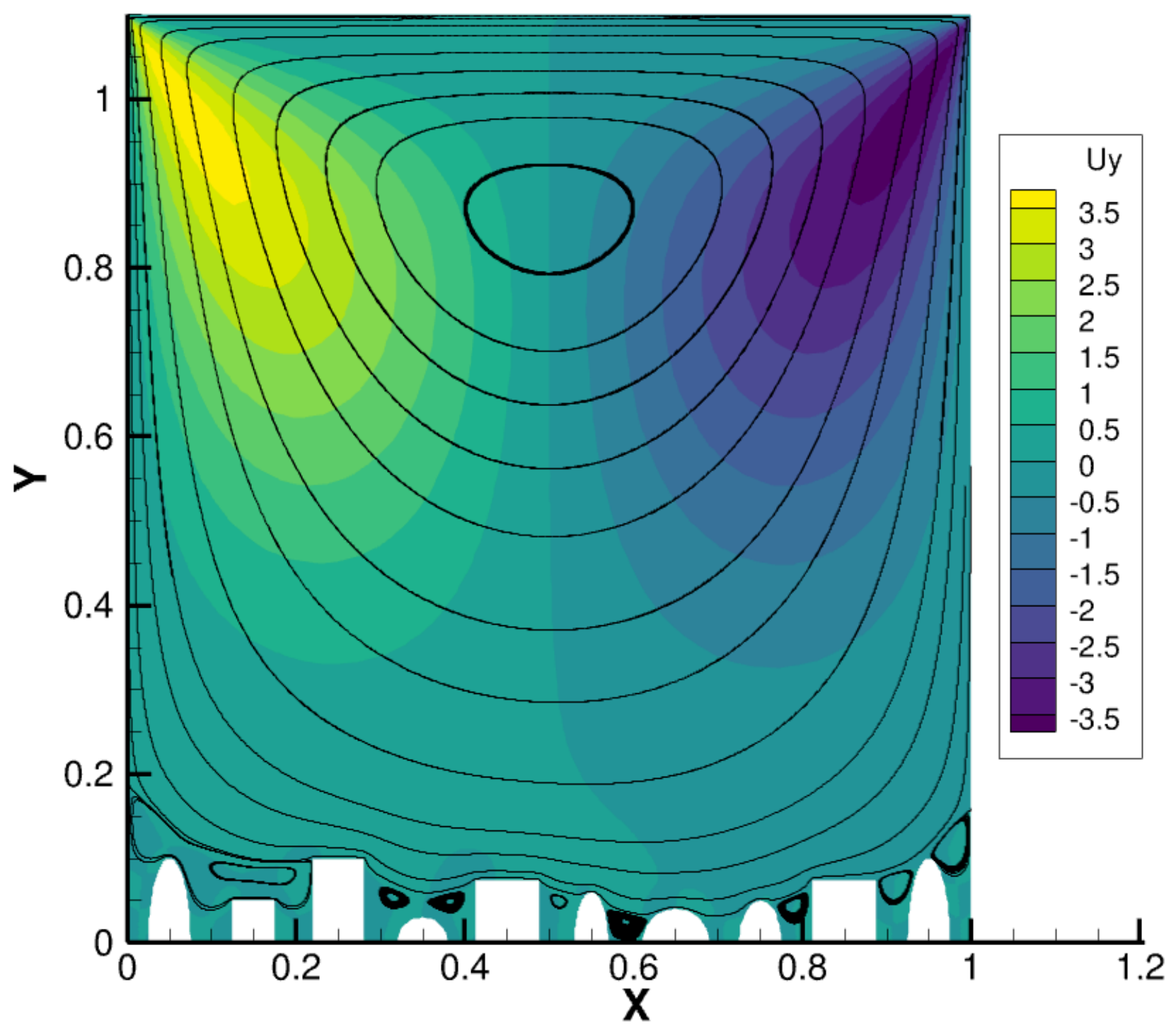}
\caption{Vertical velocity contours and streamines within the cavity with random rough elements.}
\label{fig:randconf}
\end{figure*}

\begin{figure*}
\subfloat[]{\includegraphics[scale=0.55]{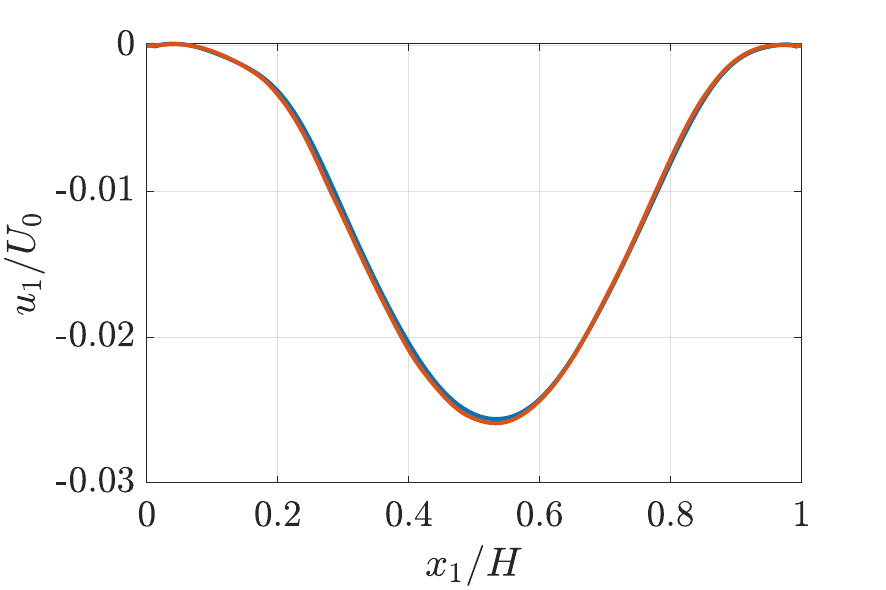}} 
\subfloat[]{\includegraphics[scale=0.55]{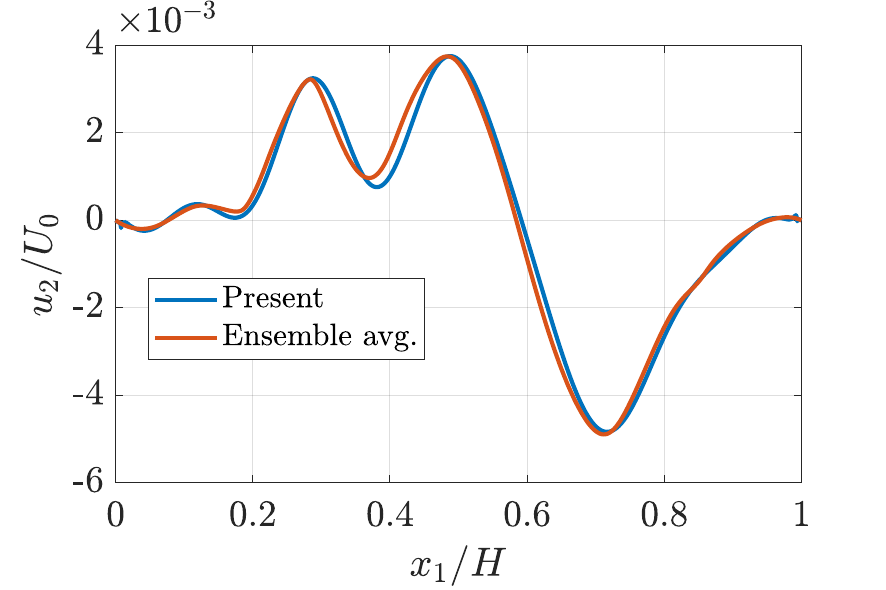}}
\caption{Comparison of interface velocities for the cavity with random rough elements (a)~tangential velocity, (b)~wall-normal velocity.}
\label{fig:randvelo}
\end{figure*}

In order to verify the above claim, we simulated a Stokes flow within a lid-driven cavity in which the bottom wall is covered with non-ordered random rough elements. The configuration is depicted in figure~\ref{fig:randconf}, which also shows contours of vertical velocity and streamlines within the cavity. Although the shape of the elements changes, we maintain the same microscopic length $l$ for each element. The comparison of ensemble averaged and macroscopic velocity components obtained from the present method is shown in figure~\ref{fig:randvelo}. The nominal interface is chosen at a vertical distance of 0.1$l$ from the tallest roughness element.

We can clearly see that the matching between both methods is good, except for a minor phase shift as shown in figure~\ref{fig:randvelo}. Notably, the present method yields a smooth variation of velocity components with only two samples, while the ensemble averaging used ten samples. The reason for the phase shift is currently unclear. As the treatment of random surfaces is beyond the scope of the present work,  it will be addressed in a future study.
\subsection{Infiltration through porous media}
In all the test cases described in the earlier sections, the  wall-normal velocity is an order of magnitude smaller than that of the tangential velocity.  In order to examine the robustness of the method, we apply it to a configuration in which both velocity components at the interface are of comparable magnitude. Infiltration flow through a porous medium~\cite{hanspal2009,carraro2015effective,eggenweiler2020,naqvi2021,strohbeck2023} provides such a test case. Eggenweiler and Rybak~\cite{eggenweiler2020} discussed the unsuitability of the classical Beavers-Joseph condition for flows oriented in an arbitrary direction with respect to the interface. This section presents the accuracy of the present method to infiltration (Stokes) flow through an anisotropic porous medium.

\begin{figure}
\includegraphics[scale=0.65]{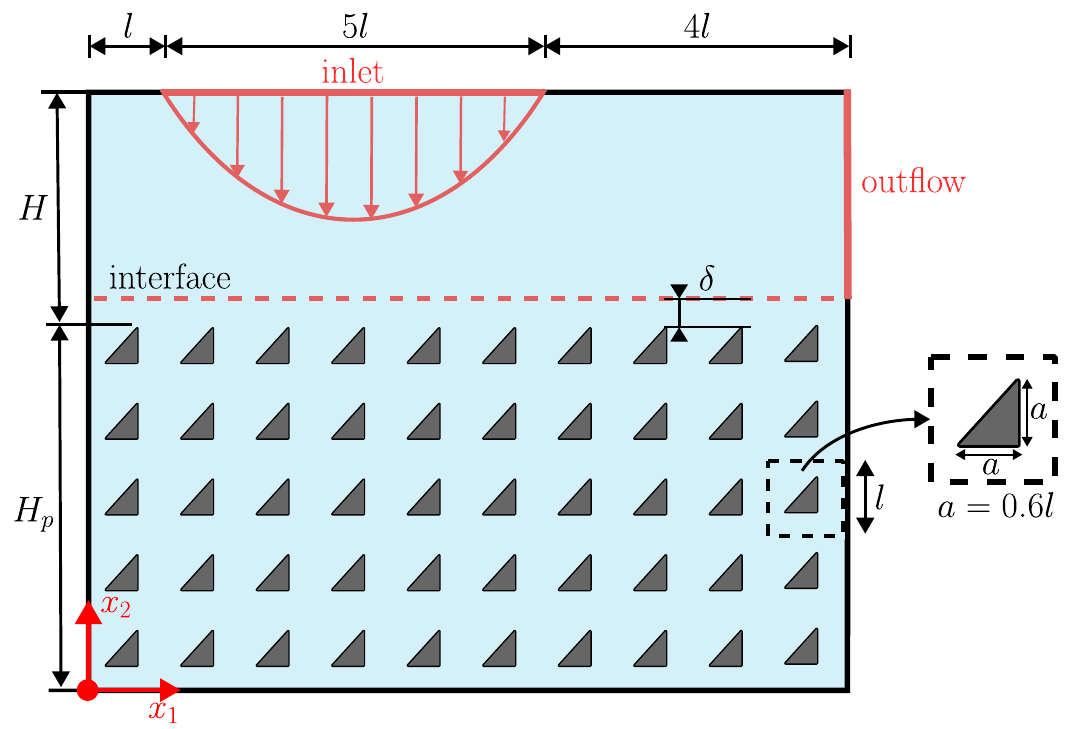}
\caption{Simulation setup for the infiltration through an anisotropic porous medium.}
\label{fig:filtrationgeom}
\end{figure}

The configuration considered,  which is a modified version of the test case studied by \citet{mohammadi2023surrogate},  is depicted in figure~\ref{fig:filtrationgeom}.  The values of the parameters are as follows: $H=2l$, and  $\delta=0.1l$.  We set $l=1$, and hence $H_p=4.8$. The fluid enters through a part of the top boundary $(1\le x_1\le 6)$ on which the the following conditions are enforced: $u_1=0$ and $u_2=-U_0\sin[0.2\pi(x_1-1)]$. Zero stress condition is applied on the outlet, which extends from the interface to the top boundary. On the rest of the domain boundaries, the homogeneous Dirichlet condition for velocity is applied.

\begin{figure*}
\subfloat[]{\includegraphics[scale=0.55]{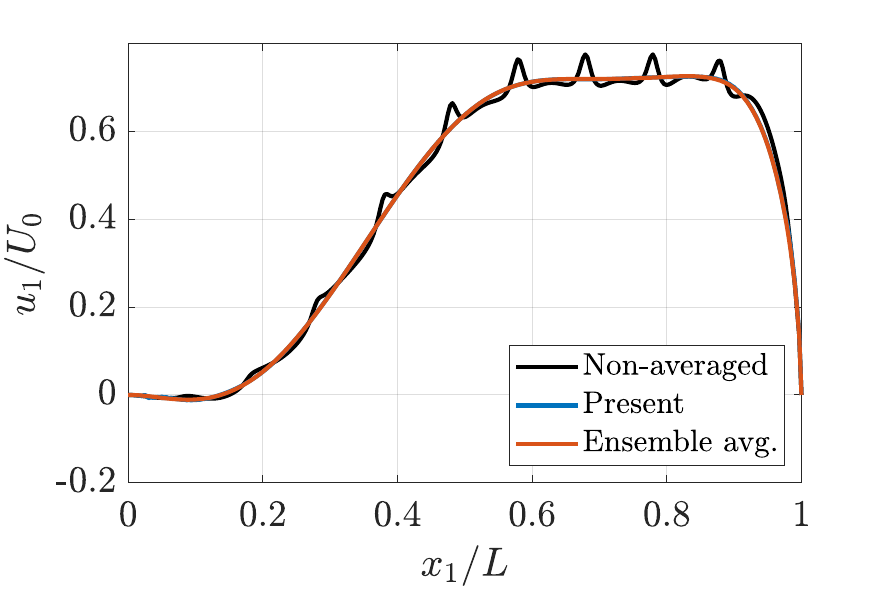}} 
\subfloat[]{\includegraphics[scale=0.55]{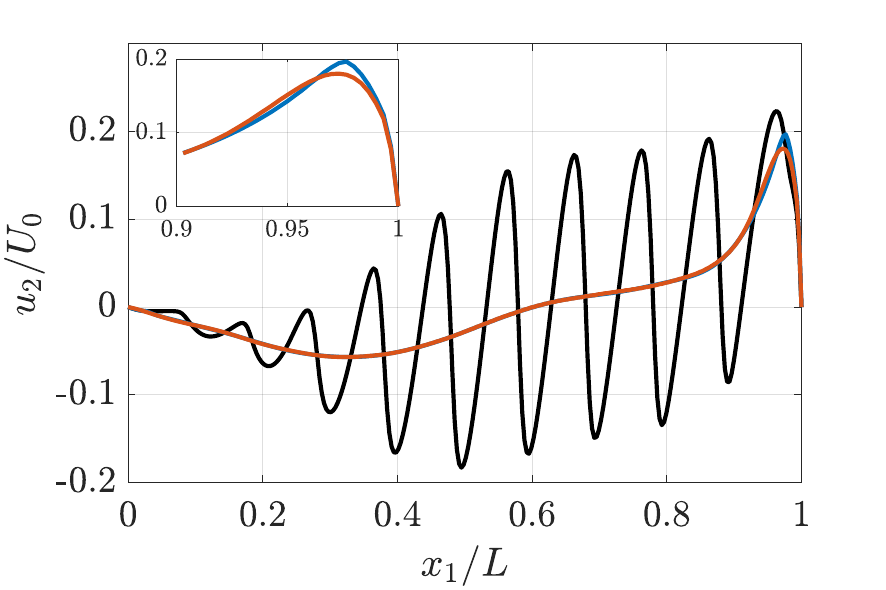}}
\caption{Velocity along the interface for the infiltration test case: (a)~tangential component (b)~wall-normal component. $L=10l$ denotes the interface length.}
\label{fig:filtrationres}
\end{figure*}

The macroscopic tangential and wall-normal velocities along the interface, and the non-averaged data are presented in figure~\ref{fig:filtrationres}.  This test case is particularly challenging than the other examples presented in this paper, due to the following reasons: (i)~the scale separation parameter is large, $\eta=l/H=0.5$; the effect of this large $\eta$ can be seen from the larger magnitude of $u_1/U_0$ and $u_2/U_0$, and (ii)~both components of macroscopic velocity involves a large gradient at the right boundary, as shown in figure~\ref{fig:filtrationres}, which might affect our smoothing algorithm. Despite these challenges, the present method is able to accurately extract macroscopic data along the interface. As can be seen from figure~\ref{fig:filtrationres}, results from the present method overlaps with those from the ensemble averaging, except for a minor overshoot in $u_2$ at the right boundary. Therefore, we can conclude that our method can produce accurate results for flows over complex surfaces, even in the presence of a large wall-normal velocity.
\section{Conclusion}
\label{sec:conc}
This paper described an efficient method to extract macroscopic interface fields from the microscopic flow fields obtained for flows over irregular surfaces using geometry-resolved simulations. The proposed method requires only two samples, even for inertia-dominated flows. Hence, it drastically reduces the amount of computational effort in getting the macroscopic data compared to the ensemble averaging technique. Results presented for different configurations proved that the method can produce accurate results for flows over rough, porous, and poroelastic media.  Moreover, it works well for parallel and infiltration flows in complex surfaces. The significance of the present work is that it enables the validation of effective models for irregular surfaces, in a computationally efficient way.  In future, we will extend this method to compute averaged results using only a single DNS data, and to handle problems with velocity boundary conditions other than Dirichlet type. Even in the current form,  to our knowledge, the present method is the only available technique applicable for flows over poroelastic surfaces exhibiting large deformation.  This feature can aid in developing effective models for such a practically important multiscale surface.

\begin{acknowledgments}
We acknowledge the financial support provided by the DST-SERB Ramanujan fellowship~(sanction order no. SB/S2/RJN-037/2018) and SERB MATRICS Grant (MTR/2021/000841).
\end{acknowledgments}

\section*{Data Availability Statement}
The source code to perform averaging using the present method, and data for rough, porous, and poroelastic walls are made available in a public repository~\cite{bitbucket}.

\nocite{*}
\bibliography{avg}
\end{document}